\UseRawInputEncoding
\documentclass[aps,prl,reprint,superscriptaddress]{revtex4-2}

\usepackage{amsmath}
\usepackage{graphicx}
\usepackage{dcolumn}
\usepackage{bm}
\usepackage[utf8]{inputenc}

\newcommand{\dmitQSL} {$\beta$'-EtMe$_3$Sb$[$Pd(dmit)$_{2}]_{2}$}

\begin{document}


\title{Quasi-one-dimensional Spin Dynamics in a Molecular Spin Liquid System}


\author{Yugo Oshima}
\email[]{yugo@riken.jp}
\affiliation{RIKEN Cluster for Pioneering Research, Wako, Saitama 351-0198, Japan.}
\author{Yasuyuki Ishii}
\affiliation{College of Engineering, Shibaura Institute of Technology, Minuma-ku, Saitama 337-8570, Japan.}
\author{Francis L. Pratt}
\affiliation{ISIS Neutron and Muon Source, STFC Rutherford Appleton Laboratory, Chilton, Oxfordshire OX11 0QX, UK}
\author{Isao Watanabe}
\affiliation{RIKEN Nishina Center, Wako, Saitama 351-0198, Japan.}
\author{Hitoshi Seo}
\affiliation{RIKEN Cluster for Pioneering Research, Wako, Saitama 351-0198, Japan.}
\affiliation{RIKEN Center for Emergence Matter Science (CEMS), 2-1 Hirosawa, Wako, Saitama 351-0198, Japan.}
\author{Takao Tsumuraya}
\affiliation{Magnesium Research Center, Kumamoto University, Kumamoto 860-8555, Japan}
\author{Tsuyoshi Miyazaki}
\affiliation{Research Center for Materials Nanoarchitectonics (MANA), National Institute for Materials Science, Tsukuba, Ibaraki 305-0044, Japan}
\author{Reizo Kato}
\affiliation{RIKEN Cluster for Pioneering Research, Wako, Saitama 351-0198, Japan.}


\date{\today}

\begin{abstract}
The molecular triangular lattice system, {\dmitQSL}, is considered as a candidate material for the quantum spin liquid (QSL) state, although ongoing debates arise from recent controversial results.
Here, the results of electron spin resonance (ESR) and muon spin relaxation ($\mu$SR) measurements on {\dmitQSL} are presented.
Both results indicate characteristic behaviors related to quasi-one-dimensional (q1D) spin dynamics, whereas the direction of anisotropy found in ESR is in contradiction with previous theories. 
We succeed in interpreting the experiments by combining density-functional theory calculations and analysis of the effective model taking into account the multi-orbital nature of the system. 
While the QSL-like origin of {\dmitQSL} was initially attributed to the magnetic frustration of the triangular lattice, it appears that the primary origin is a 1D spin liquid resulting from the dimensional reduction effect. 
\end{abstract}


\maketitle


Ground states of magnetically frustrated triangular lattice systems, where antiferromagnetically coupled $S$=1/2 spins are positioned on each lattice site, have posed a long-standing challenge in condensed matter physics since Anderson 
proposed the resonating-valence-bond (RVB) state for the $S$=1/2 Heisenberg model on a uniform triangular lattice \cite{Anderson, LeePA, BalentsNature, Lacroix, Diep, Savary, Zhou}.
Although confirmation of the quantum spin liquid (QSL) state has remained elusive, 
several molecular materials with $S$=1/2 triangular lattices have garnered significant attention \cite{Shimizu, SatoshiNatPhys, MinoruNatPhys, PrattNature, Isono, ItouPRB, Tamura, KanodaKato}.
One notable QSL candidate is the dimer-Mott insulator \dmitQSL, where dmit, Et and Me are 1,3-dithiole-2-thione-4,5-dithiolate, ethyl and methyl, respectively \cite{ItouPRB, Tamura, KanodaKato, KatoBulletin}.

\begin{figure}
\includegraphics[width=8cm]{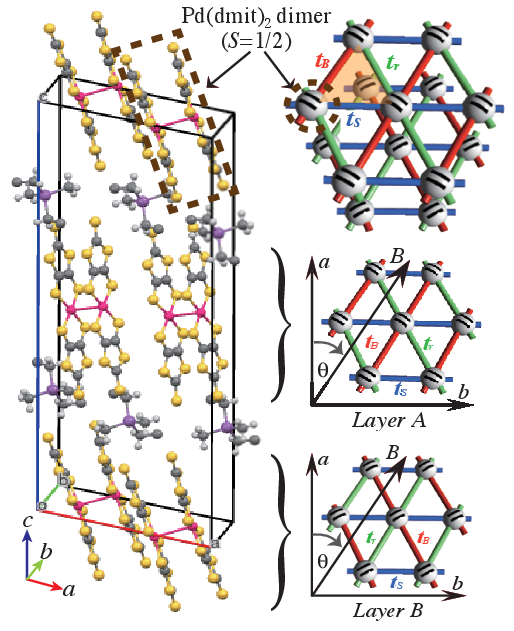}%
\caption{\label{fig:crystalStructure} (left) Crystal structure of \dmitQSL. Two crystallographically equivalent Pd(dmit)$_2$ layers with different dimer stacking directions exist in the unit cell (Layer A and B), and the Pd(dmit)$_2$ dimer with $S$=1/2 forms a triangular lattice in each layer. 
 (right) Schematic drawings of the triangular lattice and its transfer integrals for layers A and B. $\theta$ is the angle of the magnetic field from the \textit{a}-axis used for ESR. 
}
\end{figure}

{\dmitQSL} consists of Pd(dmit)$_2$ anions and a monovalent countercation EtMe$_3$Sb$^+$.
The anions form strongly dimerized $[$Pd(dmit)$_2$]$_{2}^{-}$, and an electron is localized on each dimer.
As shown in Fig. \ref{fig:crystalStructure}, the crystal structure of {\dmitQSL} has two crystallographically equivalent Pd(dmit)$_2$ layers with different dimer stacking direction (layer A and B). Layers A and B are connected by the glide plane symmetry. 
The Pd(dmit)$_2$ dimers form a $S$=1/2 triangular lattice on each layer, where the transfer integral along the dimer's stacking direction is denoted as $t_B$, the side-by-side direction as $t_S$, and the diagonal direction as $t_r$ (Fig. \ref{fig:crystalStructure}). 
The calculated transfer integrals between the dimers suggest a nearly isosceles or scalene triangular arrangement in the magnetic geometry (see Table \ref{transferTable}) \cite{Ueda, NakamuraPRB, ScrivenPRL, Tsumuraya, YoshimiPRR, IdoQM}.
{\dmitQSL} shows no magnetic long-range order down to approximately 40 mK, which is several orders of magnitude lower than its exchange interaction \cite{ItouNatPhys, MinoruScience, SatoshiNatComm, WatanabeNatComm, Poirier, Yamataka, Ni, BourgeoisHope, MinoruJPSJ, MinoruPRB, NomotoPRB}.
Precise tuning of the transfer integrals using mixed countercations further reveals the presence of a QSL `phase' in the system \cite{UedaPRB}.


However, the nature of its QSL-like ground state is so far controversial.  
The specific heat and the first report of thermal conductivity measurements both show linear temperature dependence, although it is an insulator \cite{MinoruScience, SatoshiNatComm}.
Such behavior is considered to be originating from the delocalized nature of the excitations and it was proposed to be owing to the existence of spinons with a Fermi surface. 
However, theoretical studies for a QSL state with a spinon Fermi surface find $T^{2/3}$ scaling rather than linear scaling \cite{Motrunich}.
Furthermore, in contradiction to the initial study of thermal conductivity, recent studies performed by three different groups
report the absence of any residual linear term $\kappa_{0}/T$ \cite{Ni, BourgeoisHope, NomotoPRB}, while Yamashita \textit{et al}. have reported that the linear term 
depends on the sample and in particular the cooling rate \cite{MinoruJPSJ, MinoruPRB, MinoruSciRep}. 
No cooling rate dependence is observed in X-ray diffraction, transport and NMR, though \cite{NoCoolingDepPaper}.
As for alternative ground states, several competing charge-order states are proposed through vibrational spectroscopy \cite{Yamataka}. 
A random-singlet state due to random intradimer charge disproportionation is proposed since a relaxor-type ferroelectricity was observed \cite{randomSingletJPSJ, randomSingletPRB, MajedPRB}.
Moreover, a recent \textit{ab-initio} calculation proposed that the QSL state of {\dmitQSL} is essentially a 1D spin-liquid, although it preserves some two-dimensionality \cite{IdoQM}.

\begin{table}
 \caption{\label{transferTable}Interdimer transfer integrals (in meV) along the three directions of the triangular lattice of {\dmitQSL} evaluated by different methods. The low-temperature crystal structure data has been used for our calculation. FP, TB and EHM stands for first-principles calculation, tight-binding and extended Hubbard models, respectively. The number of bands included in the analysis is indicated. }
 \begin{ruledtabular}
\begin{tabular}{ccccc}
 Methods & $t_B$ & $t_S$ & $t_r$ & Ref.\\
 \hline
Extended H\"uckel  & 34 & 33 & 26 & \cite{Ueda}\\
FP + TB (6-band) & 54 & 45 & 40 & \cite{NakamuraPRB}\\
FP + TB (2-band) & 49 & 45 & 37 & \cite{ScrivenPRL}\\
FP + TB (2-band)& 55 & 47 &39 & \cite{Tsumuraya}\\
FP + TB (2-band) & 57 & 45 & 40 & \cite{YoshimiPRR}\\
FP + TB (2-band)& 57 & 45 &40 & \cite{IdoQM}\\
FP + TB (8-band) + EHM   & 31 & 28 & 36 & present study\\
 \end{tabular}
 \end{ruledtabular}
 \end{table}

In this Letter, we present a different experimental approach for studying the ground state of {\dmitQSL}~ using electron spin resonance (ESR) 
and muon spin rotation ($\mu$SR). 
Both results exhibit characteristic features of a quasi-one-dimensional (q1D) spin dynamics, with the fastest propagation direction for spin dynamics being along $t_r$, which corresponds to the direction of weakest magnetic coupling, as indicated by previous calculations \cite{Ueda, NakamuraPRB, ScrivenPRL, Tsumuraya, YoshimiPRR, IdoQM}.
By extending the theoretical analysis to include the multi-orbital nature of the system, we show a renewed picture of the magnetic anisotropy, finding good agreement with the experimental results.  
Our finding suggests the QSL ground state of {\dmitQSL} is another example of `dimensional reduction' induced by frustration and quantum fluctuations.  

Single crystals of {\dmitQSL} were prepared using the method described in Ref. \cite{BourgeoisHope}.
ESR measurements were carried out with a conventional X-band ESR spectrometer ($\sim$ 9.1 GHz), using a single crystal (approximately 1 $\times$ 1 $\times$ 0.05 mm$^3$) mounted on a quartz rod to allow rotation in the \textit{ab}-plane.
The $\mu$SR experiments were performed at the RIKEN-RAL muon facility in the UK and the S$\mu$S facility in Switzerland.
Randomly oriented crystals, with a total weight of 100 mg, were wrapped in a packet of 12.5 $\mu$m silver foil and attached to the sample plate of a helium dilution refrigerator.
Further details of the experimental setups and theoretical calculations can be found in the Supporting Information (SI) \cite{SI}. 
\\

\begin{figure}
\includegraphics[width=7cm]{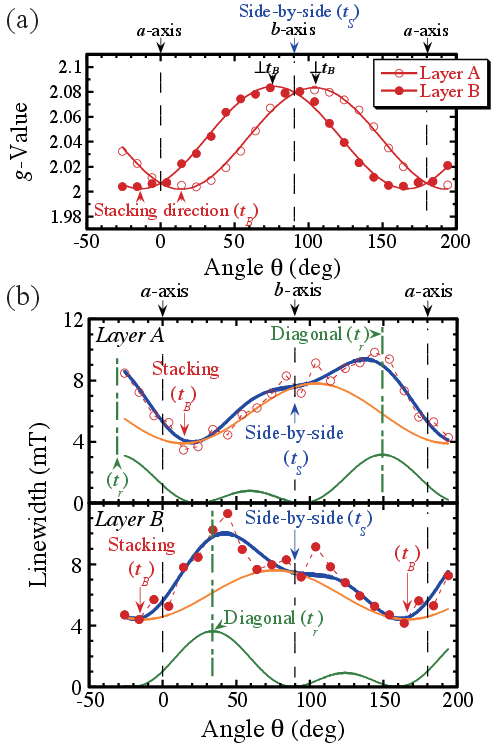}%
\caption{\label{fig:ESRAngDep} The angular dependence of (a) the $g$-value and (b) the ESR linewidth of \dmitQSL~ at 4.7 K 
where the magnetic field is rotated within the $ab$-plane ($a$-axis is at $\theta$=0$^\circ $). 
Two ESR signals from layers A and B are presented as red open and solid circles, respectively. 
The angular dependence of the linewidth is fitted with the sum (thick blue curve) of 
$1+\cos^{2} \left( \theta - \theta_{g_{\max}} \right)$ and 
$( 3 \cos^{2} \left( \theta-\theta_{q1D}\right)-1)^{2}$ terms, 
presented as orange and green solid curves, respectively. 
$\theta_{g\max}$ is 104$^\circ$ and 76$^\circ $ for layers A and B, respectively, 
which is in good agreement with the angle dependence of the $g$-tensors. 
$\theta_{q1D}$ is found to be 149$^\circ $ and 34$^\circ $ for layers A and B, respectively, 
which approximately corresponds to the diagonal direction $t_r$ of the triangular lattice in each layer.
}
\end{figure}

Two distinct ESR signals are observed when the magnetic field is rotated within the $ab$-plane of the triangular lattice of the {\dmitQSL}~ salt. 
The observed ESR spectra are presented in Fig. S2 \cite{SI}.
The angular dependence of the $g$-values, obtained from the ESR signals, 
shows two almost identical components of the $g$-tensor with a shift between them of about 30$^\circ $ (Fig. \ref{fig:ESRAngDep}(a)). 
This $g$-tensor shift coincides with the difference in the stacking direction of the Pd(dmit)$_2$ dimers between adjacent layers, as shown in Fig. \ref{fig:crystalStructure}. 
Hence, the principal axes of the $g$-tensor are related to the orientation of the Pd(dmit)$_2$ dimers ($S$=1/2), 
and the minimum and the maximum of the $g$-value are observed when the magnetic field is respectively applied 
parallel and perpendicular to the stacking direction of the Pd(dmit)$_2$ dimers. 
The perpendicular direction is close to the $b$-axis side-by-side direction of the dimer arrangement. 
From the comparison of the $g$-values with the crystal axes, the two ESR signals can be assigned 
to layer A and layer B of the Pd(dmit)$_2$ layers, shown as open and solid red circles in Fig. \ref{fig:ESRAngDep}(a), respectively. 
These results show that the ESR origin is purely from the spins on the triangular lattice, and extrinsic effects from impurities are ruled out.

It is well known that if a finite exchange interaction exists between two spins with different $g$-tensors, 
the two independent ESR absorption lines merge into a single absorption line in a process known as exchange narrowing \cite{AndersonJPSJ, BenciniGatteschi}.
The exchange interaction $J$ can be roughly estimated from the relation $2J \sim | \Delta g | \mu_{B} B$, where $\Delta g$ is the difference in the $g$-values when the amalgamation of ESR lines occurs. \cite{OkudaDate, AndersonJPSJ}
From Fig. \ref{fig:ESRAngDep}(a), we observe that the amalgamation occurs where $\Delta g$ = 0.005, leading to an estimated interlayer exchange interaction of approximately 0.54 mK. 
This small interlayer interaction suggests that the magnetic network of {\dmitQSL} is highly 2D, which is consistent with the absence of long-range magnetic order in this system.

The in-plane angular dependence of the ESR linewidth at 4.7 K obtained from ESR of layers A and B are presented, respectively, as open and solid circles in Fig. \ref{fig:ESRAngDep}(b).
The linewidth shows an unconventional angular dependence, not previously seen in the case of a low-dimensional system 
or an inorganic triangular lattice \cite{BenciniGatteschi, Fayzullin}.
We find that this unusual angular dependence is well-fitted by a sum of $(1+\cos ^{2}\theta)$ and $(3\cos ^{2}\theta-1)^2$ terms, 
which are presented respectively as orange and green solid curves in Fig. \ref{fig:ESRAngDep}(b). 
The former angular dependence is a typical one for angle-dependent ESR and originates from the contributions of magnetic anisotropies 
to the $g$-values and dipolar or hyperfine interactions. 
For each layer, the extrema of the $(1+\cos ^{2}\theta)$ term coincide with those of the $g$-values ($t_B$ and $\perp t_B$, respectively). 
The $(3\cos ^{2}\theta-1)^2$ angular term originates from q1D spin diffusion, 
which is commonly observed in ESR studies of low dimensional spin systems \cite{BenciniGatteschi, Hennessy, Cheung, Takahashi, Tanaka}.
For q1D spin diffusion, the direction showing the maximum of the linewidth corresponds to the diffusion direction. 
Our results show that the diffusive direction is along the diagonal direction of the triangular lattice ($t_r$) for both layers A and B. 
Despite our system consisting of a 2D triangular magnetic network, the ESR results indicate a q1D spin dynamics along $t_r$.

Next, we present our $\mu$SR results for \dmitQSL.
Neither a spontaneous precession signal nor a divergence of the muon depolarization rate $\lambda$ 
was observed down to low temperatures under zero-field conditions (ZF-$\mu$SR, Fig. S3(a) in SI) \cite{SI}.
These results show that there is no sign of long-range order down to 28 mK, 
in good agreement with the specific heat and thermal conductivity measurements \cite{MinoruScience, SatoshiNatComm, Ni, BourgeoisHope, MinoruJPSJ, MinoruPRB}.
Moreover, no excitation gap is found from the field dependence of the transverse-field muon spin rotation (TF-$\mu$SR) measurements (Fig. S4).

\begin{figure}
\includegraphics[width=8.4cm]{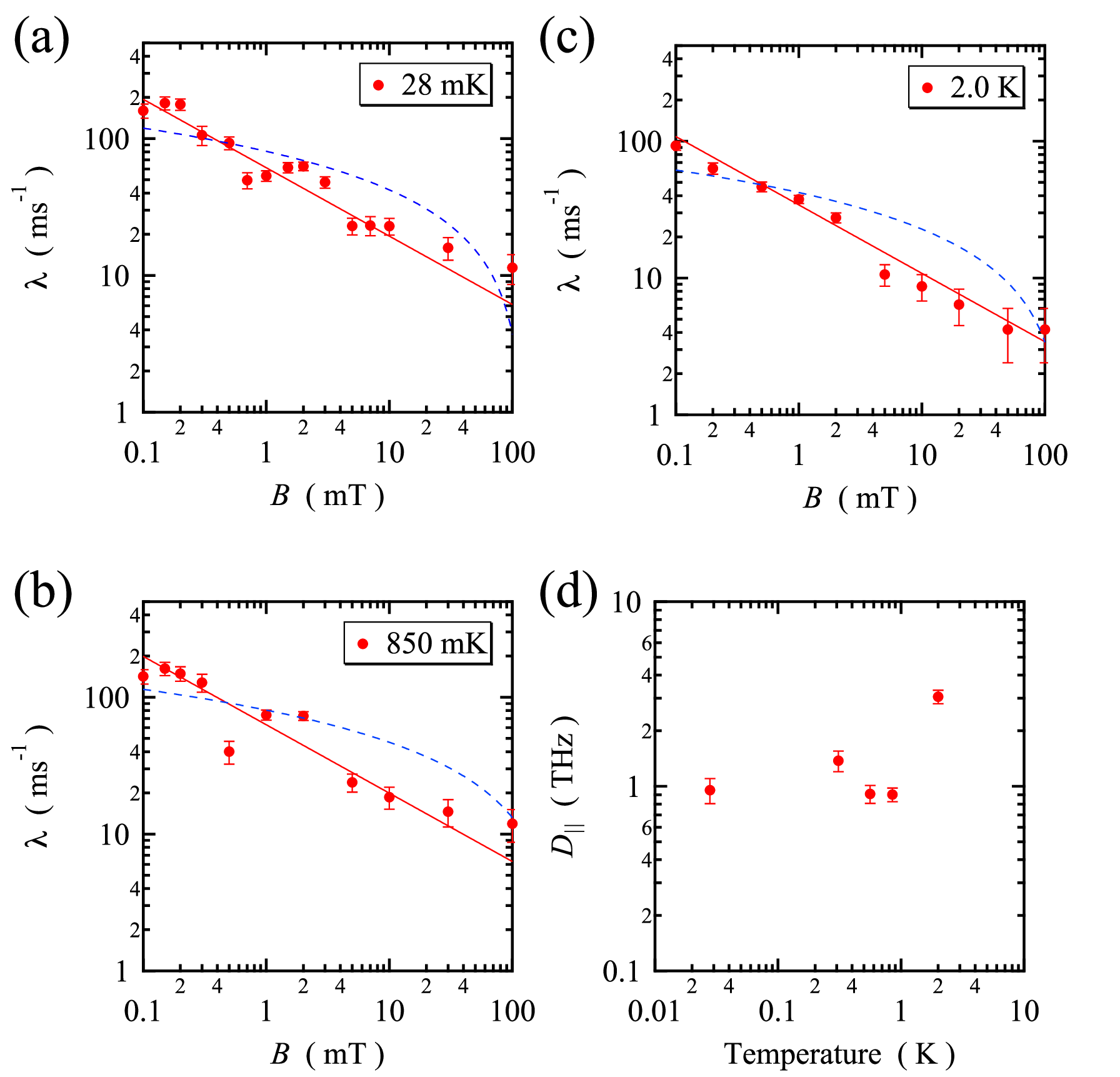}%
\caption{\label{fig:relaxationRatemSR} 
Magnetic field dependence of the relaxation rate $\lambda$ of {\dmitQSL} at (a) 28 mK, (b) 0.85 K, and (c) 2.0~K. 
The red solid lines and the blue dashed lines are the best fits from the q1D model and the 2D diffusive model, respectively. 
(d) Temperature dependence of intra-chain diffusion rate $D_\parallel$.
}
\end{figure}

To gain information from the spin dynamics, the field dependence of $\lambda$ has been studied. 
Longitudinal-fields were applied along the muon-spin polarization direction (LF-$\mu$SR) and 
the field-dependent muon-spin depolarization rate $\lambda$(B) was measured at various temperatures (Figs. \ref{fig:relaxationRatemSR}(a)-(c)). 
Each one shows a $B^{-0.5}$ dependence over a wide field range of 0.1 $< B <$ 100 mT (red line in Figs. \ref{fig:relaxationRatemSR}(a)-(c)) 
at temperatures below 2 K. 
This characteristic behavior reflects a 1D spin diffusion, which is fully in accordance with the ESR results. 
Note that our results cannot be fitted with a 2D diffusive model (blue dashed lines).  
From these field dependences, we could obtain diffusion rates $D_{\parallel}$ using the expression
\begin{eqnarray}
\lambda(B)=\frac{A^2}{4}(2D_{\parallel} \gamma_{e} B)^{-1/2}
\end{eqnarray}
where $A$ is a scalar hyperfine coupling constant and $\gamma_e$ is a gyromagnetic ratio of electron. 
Its derivation can be found in the SI \cite{SI}.
Our DFT calculations for muon additions to Pd(dmit)$_2$ using Gaussian16 \cite{Gaussian} show the lowest energy when the muons were added to the end S site of Pd(dmit)$_2$ molecule, with the corresponding experimental value of 71 MHz for $A$ \cite{SI}.
This value is comparable with the value of 82 MHz found for muonium addition at the end of the electron acceptor molecule TCNQ \cite{PrattMRC}.
Using the obtained coupling constant, the diffusion rate $D_{\parallel}$ is estimated as being of the order of $10^{12}$ s$^{-1}$.
The temperature dependence of $D_{||}$ is shown in Fig. \ref{fig:relaxationRatemSR}(d). 
We can also evaluate the degree of one-dimensionality from the data, 
giving a lower limit estimate of the ratio of intra-chain to inter-chain diffusion rate as $D_{\parallel}/D_{\perp}>10^4$ \cite{SI}. 
This implies that the diffusion is highly 1D 
despite the three transfer integrals around the triangular unit having rather similar values, according to calculation.

\begin{figure}
\includegraphics[width=7.4cm]{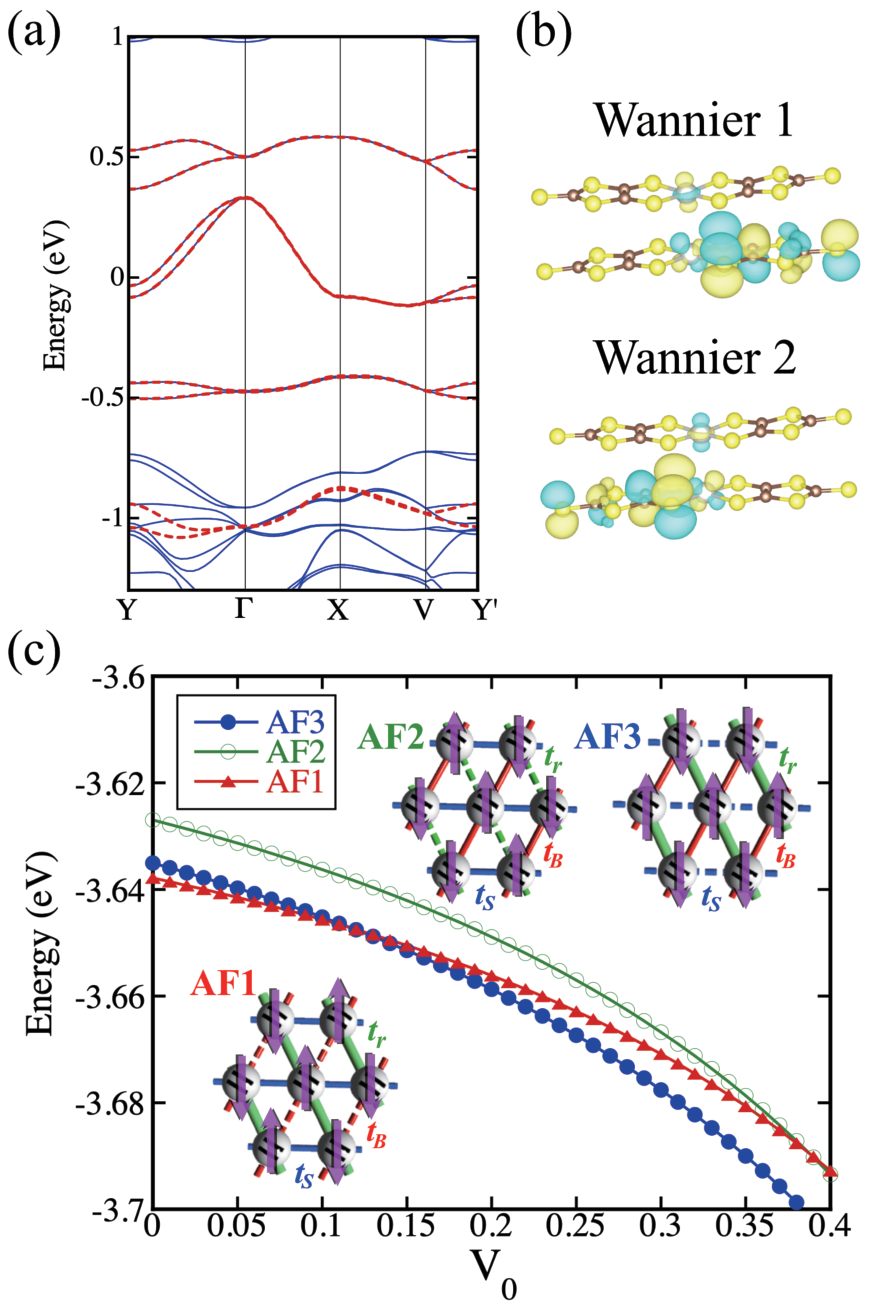}%
\caption{\label{fig:calculation} 
(a) Band structure of {\dmitQSL} (blue solid curves), and the TB bands based on the MLWFs (red broken curves). 
(b) Two independent MLWF in the Pd(dmit)$_2$ dimer, drawn using \texttt{VESTA}~\cite{VESTA}. 
(c) Mean-field energies of different AF patterns varying the intersite Coulomb energies scaled by $V_0$. The onsite Coulomb repulsion is set to $U$=1.0 eV. The AF alignments for each stable pattern are shown in the inset.
}
\end{figure}

The $\mu$SR and ESR results reveals a gapless ground state with a q1D spin dynamics.
This might suggest the QSL-like ground state of {\dmitQSL} is essentially a 1D spin-liquid.
However, the diffusive direction is found to be along $t_r$, which is the smallest transfer integral from previous theoretical estimations as shown in Table I \cite{Ueda, NakamuraPRB, ScrivenPRL, Tsumuraya, YoshimiPRR, IdoQM}.

Following these results, we have reanalyzed its electronic structure (see Ref. \cite{SI} for details). 
Using first-principles calculations based on the density functional theory (DFT), we derive the maximally-localized Wannier functions (MLWFs) from 8 bands near the Fermi level. 
Here, following recent studies \cite{MisawaPRR,YoshimiPRR}, plane-wave DFT calculation within the generalized gradient approximation \cite{PerdewPRL} was performed with the QUANTUM ESPRESSO code \cite{Giannozzi} using norm-conserving pseudopotentials \cite{HamannPRB,Schlipf}. 
MLWFs were generated using the Wannier90 package \cite{MarzariPRB, SouzaPRB}, by setting a Wannier center at each dmit ligand. 
Note that these MLWFs form not only the highest occupied molecular orbital (HOMO) but also the lowest occupied molecular orbital (LUMO), and eight bands are formed from them since there are four Pd(dmit)$_2$ molecules in the unit cell.
In Fig. \ref{fig:calculation}(a), the DFT band structure is shown together with the tight-binding (TB) bands based on the MLWFs, whose spatial forms are shown in Fig. \ref{fig:calculation}(b). 
The MLWFs are distributed on either side of the molecule, that show similarity to the wave functions introduced and called fragment molecular orbital (fMO) in Ref. \cite{SeoJPSJ}. 

Based on the derived TB parameters, we investigate the electron correlation effect by the extended Hubbard model and compare the mean-field energies of different antiferromagnetic (AF) patterns. 
Fig. \ref{fig:calculation}(c) shows the result for onsite Coulomb repulsion $U=1.0$ eV, and varying the intersite Coulomb energies scaled by $V_0$. 
One can see that AF1 and AF3 states are competing in energy, in which both patterns are AF along the $t_r$ direction. 
This suggests that this direction indeed shows the strongest magnetic interaction. 

We can evaluate the effective interdimer transfer integrals based on these calculations, following Ref. \cite{SeoJPSJ}. 
The wave function where $S=1/2$ is localized is described by the linear combination of the fMOs, and their weights can be adopted from the mean-field solution which is about 1:3 within the molecule, consistent with the NMR measurement \cite{FujiyamaPRL}. 
Typical parameters result in the values listed in Table I: Intriguingly, $t_r$ becomes the largest. 
This difference from previous studies comes from the multi-orbital nature; most studies have focused on the half-filled valence bands based on HOMO.  
However, the HOMO-LUMO levels in the isolated Pd(dmit)$_2$ molecule are close in energy with a separation of about 0.5 eV, while the Coulomb energy is of the same order \cite{Ueda, Tsumuraya, IdoQM, MiyazakiPRB}. 
It is then natural to consider the multiple orbitals, and we have shown its significance in this study. 

The competition and the fluctuation of the AF1 and AF3 states might suggest that the dimensional reduction effect takes place in the frustrated triangular lattice.
In a similar manner to the triangular lattice system Cs$_2$CuCl$_4$, the magnetic frustration significantly reduces the interchain correlations in the ground state, and 1D physics similar to the spin-chain system can appear \cite{ColdeaPRL, BalentsNature, Kohno, DimensionalReduction}.
Therefore, we conclude that the ground state of {\dmitQSL} might be a 1D spin liquid rather than a 2D QSL state of the triangular lattice. 

Cs$_2$CuCl$_4$ exhibits dimensional reduction within an isosceles triangular magnetic lattice where \textit{J'/J} $\sim$ 0.4, and a small interlayer coupling stabilizes the 3D magnetic order below $T_N$=0.6 K \cite{ColdeaPRL, BalentsNature, Kohno, DimensionalReduction}.
In contrast, {\dmitQSL} shows dimensional reduction even with the larger ratio $J'/J \sim 0.7$ (from the ratio of $t^2$ values), without any sign of long-range order.
Such significant dimensional reduction and the lack of long-range order cannot be explained by a simple \textit{S=}1/2 Heisenberg model, and other factors, such as charge and orbital degrees of freedom and the infinitesimal interlayer interaction, which causes AF instability, should be taken into account.
Let us note that the range of $J'/J$  where dimensional reduction and gapless excitation are theoretically observed remains controversial.
For example, RVB type theories suggest $0 \leq J'/J \leq 0.25$ in Ref. \cite{Hayashi} and $0 \leq J'/J \leq 0.65$ in Ref. \cite{Yunoki}; the latter may give the upper bound. 
Further development of such theoretical approach might be able to account for the dimensional reduction we observe at the larger value of $J'/J \sim 0.7$.
It is possible that the cation's orientational disorder in {\dmitQSL} also contributes to the AF instability \cite{KatoBulletin}.
However, the cation disorder does not seem to affect the magnetic network, as the diffusion anisotropy ratio remains highly 1D ($D_{\parallel}/D_{\perp}>10^4$).
Experimentally, the absence of long-range magnetic order in this system appears to be primarily determined by the infinitesimal interlayer interactions.

\begin{acknowledgments}
Muon site calculations were carried out using the STFC SCARF compute cluster. 
Part of this work was carried out at the ISIS Neutron and Muon Source, STFC Rutherford Appleton Laboratory, U.K.
This work was partially supported by JSPS KAKENHI Grant Numbers JP20H04463 (IW), 23H01129 (HS), 23H04047 (HS), 23K03333 (HS).
\end{acknowledgments}

%


\begin{thebibliography}{67}%
\makeatletter
\providecommand \@ifxundefined [1]{%
 \@ifx{#1\undefined}
}%
\providecommand \@ifnum [1]{%
 \ifnum #1\expandafter \@firstoftwo
 \else \expandafter \@secondoftwo
 \fi
}%
\providecommand \@ifx [1]{%
 \ifx #1\expandafter \@firstoftwo
 \else \expandafter \@secondoftwo
 \fi
}%
\providecommand \natexlab [1]{#1}%
\providecommand \enquote  [1]{``#1''}%
\providecommand \bibnamefont  [1]{#1}%
\providecommand \bibfnamefont [1]{#1}%
\providecommand \citenamefont [1]{#1}%
\providecommand \href@noop [0]{\@secondoftwo}%
\providecommand \href [0]{\begingroup \@sanitize@url \@href}%
\providecommand \@href[1]{\@@startlink{#1}\@@href}%
\providecommand \@@href[1]{\endgroup#1\@@endlink}%
\providecommand \@sanitize@url [0]{\catcode `\\12\catcode `\$12\catcode
  `\&12\catcode `\#12\catcode `\^12\catcode `\_12\catcode `\%12\relax}%
\providecommand \@@startlink[1]{}%
\providecommand \@@endlink[0]{}%
\providecommand \url  [0]{\begingroup\@sanitize@url \@url }%
\providecommand \@url [1]{\endgroup\@href {#1}{\urlprefix }}%
\providecommand \urlprefix  [0]{URL }%
\providecommand \Eprint [0]{\href }%
\providecommand \doibase [0]{https://doi.org/}%
\providecommand \selectlanguage [0]{\@gobble}%
\providecommand \bibinfo  [0]{\@secondoftwo}%
\providecommand \bibfield  [0]{\@secondoftwo}%
\providecommand \translation [1]{[#1]}%
\providecommand \BibitemOpen [0]{}%
\providecommand \bibitemStop [0]{}%
\providecommand \bibitemNoStop [0]{.\EOS\space}%
\providecommand \EOS [0]{\spacefactor3000\relax}%
\providecommand \BibitemShut  [1]{\csname bibitem#1\endcsname}%
\let\auto@bib@innerbib\@empty
\bibitem [{\citenamefont {Anderson}(1973)}]{Anderson}%
  \BibitemOpen
  \bibfield  {author} {\bibinfo {author} {\bibfnamefont {P.~W.}\ \bibnamefont
  {Anderson}},\ }\href
  {http://www.sciencedirect.com/science/article/pii/0025540873901670}
  {\bibfield  {journal} {\bibinfo  {journal} {Mater. Res. Bull.}\ }\textbf
  {\bibinfo {volume} {8}},\ \bibinfo {pages} {153} (\bibinfo {year}
  {1973})}\BibitemShut {NoStop}%
\bibitem [{\citenamefont {Lee}(2008)}]{LeePA}%
  \BibitemOpen
  \bibfield  {author} {\bibinfo {author} {\bibfnamefont {P.~A.}\ \bibnamefont
  {Lee}},\ }\href {http://scholar.google.com/scholar?q=An End to the Drought of
  Quantum Spin Liquid&btnG=&hl=en&num=20&as_sdt=0%2C22} {\bibfield  {journal}
  {\bibinfo  {journal} {Science}\ }\textbf {\bibinfo {volume} {321}},\ \bibinfo
  {pages} {1306} (\bibinfo {year} {2008})}\BibitemShut {NoStop}%
\bibitem [{\citenamefont {Balents}(2010)}]{BalentsNature}%
  \BibitemOpen
  \bibfield  {author} {\bibinfo {author} {\bibfnamefont {L.}~\bibnamefont
  {Balents}},\ }\href {http://dx.doi.org/10.1038/nature08917} {\bibfield
  {journal} {\bibinfo  {journal} {Nature}\ }\textbf {\bibinfo {volume} {464}},\
  \bibinfo {pages} {199} (\bibinfo {year} {2010})}\BibitemShut {NoStop}%
\bibitem [{\citenamefont {Lacroix}\ \emph {et~al.}(2011)\citenamefont
  {Lacroix}, \citenamefont {Mendels},\ and\ \citenamefont {Mila}}]{Lacroix}%
  \BibitemOpen
  \bibfield  {author} {\bibinfo {author} {\bibfnamefont {C.}~\bibnamefont
  {Lacroix}}, \bibinfo {author} {\bibfnamefont {P.}~\bibnamefont {Mendels}},\
  and\ \bibinfo {author} {\bibfnamefont {F.}~\bibnamefont {Mila}},\ }\href@noop
  {} {\emph {\bibinfo {title} {(Eds.) Introduction to frustrated magnetism:
  Materials, Experiments, Theory}}},\ Springer series in solid-state sciences\
  (\bibinfo  {publisher} {Springer},\ \bibinfo {address} {Berlin},\ \bibinfo
  {year} {2011})\BibitemShut {NoStop}%
\bibitem [{\citenamefont {Diep}(2013)}]{Diep}%
  \BibitemOpen
  \bibfield  {author} {\bibinfo {author} {\bibfnamefont {H.~T.}\ \bibnamefont
  {Diep}},\ }\href@noop {} {\emph {\bibinfo {title} {Frustrated Spin System 2nd
  Edition}}}\ (\bibinfo  {publisher} {World Scientific},\ \bibinfo {year}
  {2013})\BibitemShut {NoStop}%
\bibitem [{\citenamefont {Savary}\ and\ \citenamefont
  {Balents}(2016)}]{Savary}%
  \BibitemOpen
  \bibfield  {author} {\bibinfo {author} {\bibfnamefont {L.}~\bibnamefont
  {Savary}}\ and\ \bibinfo {author} {\bibfnamefont {L.}~\bibnamefont
  {Balents}},\ }\href {http://dx.doi.org/10.1088/0034-4885/80/1/016502
  https://iopscience.iop.org/article/10.1088/0034-4885/80/1/016502} {\bibfield
  {journal} {\bibinfo  {journal} {Rep. Prog. Phys.}\ }\textbf {\bibinfo
  {volume} {80}},\ \bibinfo {pages} {016502} (\bibinfo {year}
  {2016})}\BibitemShut {NoStop}%
\bibitem [{\citenamefont {Zhou}\ \emph {et~al.}(2017)\citenamefont {Zhou},
  \citenamefont {Kanoda},\ and\ \citenamefont {Ng}}]{Zhou}%
  \BibitemOpen
  \bibfield  {author} {\bibinfo {author} {\bibfnamefont {Y.}~\bibnamefont
  {Zhou}}, \bibinfo {author} {\bibfnamefont {K.}~\bibnamefont {Kanoda}},\ and\
  \bibinfo {author} {\bibfnamefont {T.~K.}\ \bibnamefont {Ng}},\ }\href {<Go to
  ISI>://WOS:000399385200001
  https://journals.aps.org/rmp/pdf/10.1103/RevModPhys.89.025003} {\bibfield
  {journal} {\bibinfo  {journal} {Rev. Mod. Phys.}\ }\textbf {\bibinfo {volume}
  {89}},\ \bibinfo {pages} {025003} (\bibinfo {year} {2017})}\BibitemShut
  {NoStop}%
\bibitem [{\citenamefont {Shimizu}\ \emph {et~al.}(2006)\citenamefont
  {Shimizu}, \citenamefont {Miyagawa}, \citenamefont {Kanoda}, \citenamefont
  {Maesato},\ and\ \citenamefont {Saito}}]{Shimizu}%
  \BibitemOpen
  \bibfield  {author} {\bibinfo {author} {\bibfnamefont {Y.}~\bibnamefont
  {Shimizu}}, \bibinfo {author} {\bibfnamefont {K.}~\bibnamefont {Miyagawa}},
  \bibinfo {author} {\bibfnamefont {K.}~\bibnamefont {Kanoda}}, \bibinfo
  {author} {\bibfnamefont {M.}~\bibnamefont {Maesato}},\ and\ \bibinfo {author}
  {\bibfnamefont {G.}~\bibnamefont {Saito}},\ }\href
  {https://journals.aps.org/prb/abstract/10.1103/PhysRevB.73.140407} {\bibfield
   {journal} {\bibinfo  {journal} {Phys. Rev. B}\ }\textbf {\bibinfo {volume}
  {73}},\ \bibinfo {pages} {140407(R)} (\bibinfo {year} {2006})}\BibitemShut
  {NoStop}%
\bibitem [{\citenamefont {Yamashita}\ \emph {et~al.}(2008)\citenamefont
  {Yamashita}, \citenamefont {Nakazawa}, \citenamefont {Oguni}, \citenamefont
  {Oshima}, \citenamefont {Nojiri}, \citenamefont {Shimizu}, \citenamefont
  {Miyagawa},\ and\ \citenamefont {Kanoda}}]{SatoshiNatPhys}%
  \BibitemOpen
  \bibfield  {author} {\bibinfo {author} {\bibfnamefont {S.}~\bibnamefont
  {Yamashita}}, \bibinfo {author} {\bibfnamefont {Y.}~\bibnamefont {Nakazawa}},
  \bibinfo {author} {\bibfnamefont {M.}~\bibnamefont {Oguni}}, \bibinfo
  {author} {\bibfnamefont {Y.}~\bibnamefont {Oshima}}, \bibinfo {author}
  {\bibfnamefont {H.}~\bibnamefont {Nojiri}}, \bibinfo {author} {\bibfnamefont
  {Y.}~\bibnamefont {Shimizu}}, \bibinfo {author} {\bibfnamefont
  {K.}~\bibnamefont {Miyagawa}},\ and\ \bibinfo {author} {\bibfnamefont
  {K.}~\bibnamefont {Kanoda}},\ }\href {http://dx.doi.org/10.1038/nphys942
  https://www.nature.com/articles/nphys942.pdf} {\bibfield  {journal} {\bibinfo
   {journal} {Nat. Phys.}\ }\textbf {\bibinfo {volume} {4}},\ \bibinfo {pages}
  {459} (\bibinfo {year} {2008})}\BibitemShut {NoStop}%
\bibitem [{\citenamefont {Yamashita}\ \emph {et~al.}(2009)\citenamefont
  {Yamashita}, \citenamefont {Nakata}, \citenamefont {Kasahara}, \citenamefont
  {Sasaki}, \citenamefont {Yoneyama}, \citenamefont {Kobayashi}, \citenamefont
  {Fujimoto}, \citenamefont {Shibauchi},\ and\ \citenamefont
  {Matsuda}}]{MinoruNatPhys}%
  \BibitemOpen
  \bibfield  {author} {\bibinfo {author} {\bibfnamefont {M.}~\bibnamefont
  {Yamashita}}, \bibinfo {author} {\bibfnamefont {N.}~\bibnamefont {Nakata}},
  \bibinfo {author} {\bibfnamefont {Y.}~\bibnamefont {Kasahara}}, \bibinfo
  {author} {\bibfnamefont {T.}~\bibnamefont {Sasaki}}, \bibinfo {author}
  {\bibfnamefont {N.}~\bibnamefont {Yoneyama}}, \bibinfo {author}
  {\bibfnamefont {N.}~\bibnamefont {Kobayashi}}, \bibinfo {author}
  {\bibfnamefont {S.}~\bibnamefont {Fujimoto}}, \bibinfo {author}
  {\bibfnamefont {T.}~\bibnamefont {Shibauchi}},\ and\ \bibinfo {author}
  {\bibfnamefont {Y.}~\bibnamefont {Matsuda}},\ }\href
  {http://dx.doi.org/10.1038/nphys1134
  https://www.nature.com/articles/nphys1134.pdf} {\bibfield  {journal}
  {\bibinfo  {journal} {Nat. Phys.}\ }\textbf {\bibinfo {volume} {5}},\
  \bibinfo {pages} {44} (\bibinfo {year} {2009})}\BibitemShut {NoStop}%
\bibitem [{\citenamefont {Pratt}\ \emph {et~al.}(2011)\citenamefont {Pratt},
  \citenamefont {Baker}, \citenamefont {Blundell}, \citenamefont {Lancaster},
  \citenamefont {Ohira-Kawamura}, \citenamefont {Baines}, \citenamefont
  {Shimizu}, \citenamefont {Kanoda}, \citenamefont {Watanabe},\ and\
  \citenamefont {Saito}}]{PrattNature}%
  \BibitemOpen
  \bibfield  {author} {\bibinfo {author} {\bibfnamefont {F.~L.}\ \bibnamefont
  {Pratt}}, \bibinfo {author} {\bibfnamefont {P.~J.}\ \bibnamefont {Baker}},
  \bibinfo {author} {\bibfnamefont {S.~J.}\ \bibnamefont {Blundell}}, \bibinfo
  {author} {\bibfnamefont {T.}~\bibnamefont {Lancaster}}, \bibinfo {author}
  {\bibfnamefont {S.}~\bibnamefont {Ohira-Kawamura}}, \bibinfo {author}
  {\bibfnamefont {C.}~\bibnamefont {Baines}}, \bibinfo {author} {\bibfnamefont
  {Y.}~\bibnamefont {Shimizu}}, \bibinfo {author} {\bibfnamefont
  {K.}~\bibnamefont {Kanoda}}, \bibinfo {author} {\bibfnamefont
  {I.}~\bibnamefont {Watanabe}},\ and\ \bibinfo {author} {\bibfnamefont
  {G.}~\bibnamefont {Saito}},\ }\href {http://dx.doi.org/10.1038/nature09910
  https://www.nature.com/articles/nature09910.pdf} {\bibfield  {journal}
  {\bibinfo  {journal} {Nature}\ }\textbf {\bibinfo {volume} {471}},\ \bibinfo
  {pages} {612} (\bibinfo {year} {2011})}\BibitemShut {NoStop}%
\bibitem [{\citenamefont {Isono}\ \emph {et~al.}(2014)\citenamefont {Isono},
  \citenamefont {Kamo}, \citenamefont {Ueda}, \citenamefont {Takahashi},
  \citenamefont {Kimata}, \citenamefont {Tajima}, \citenamefont {Tsuchiya},
  \citenamefont {Terashima}, \citenamefont {Uji},\ and\ \citenamefont
  {Mori}}]{Isono}%
  \BibitemOpen
  \bibfield  {author} {\bibinfo {author} {\bibfnamefont {T.}~\bibnamefont
  {Isono}}, \bibinfo {author} {\bibfnamefont {H.}~\bibnamefont {Kamo}},
  \bibinfo {author} {\bibfnamefont {A.}~\bibnamefont {Ueda}}, \bibinfo {author}
  {\bibfnamefont {K.}~\bibnamefont {Takahashi}}, \bibinfo {author}
  {\bibfnamefont {M.}~\bibnamefont {Kimata}}, \bibinfo {author} {\bibfnamefont
  {H.}~\bibnamefont {Tajima}}, \bibinfo {author} {\bibfnamefont
  {S.}~\bibnamefont {Tsuchiya}}, \bibinfo {author} {\bibfnamefont
  {T.}~\bibnamefont {Terashima}}, \bibinfo {author} {\bibfnamefont
  {S.}~\bibnamefont {Uji}},\ and\ \bibinfo {author} {\bibfnamefont
  {H.}~\bibnamefont {Mori}},\ }\href
  {https://journals.aps.org/prl/abstract/10.1103/PhysRevLett.112.177201}
  {\bibfield  {journal} {\bibinfo  {journal} {Phys. Rev. Lett.}\ }\textbf
  {\bibinfo {volume} {112}},\ \bibinfo {pages} {177201} (\bibinfo {year}
  {2014})}\BibitemShut {NoStop}%
\bibitem [{\citenamefont {Itou}\ \emph {et~al.}(2008)\citenamefont {Itou},
  \citenamefont {Oyamada}, \citenamefont {Maegawa}, \citenamefont {Tamura},\
  and\ \citenamefont {Kato}}]{ItouPRB}%
  \BibitemOpen
  \bibfield  {author} {\bibinfo {author} {\bibfnamefont {T.}~\bibnamefont
  {Itou}}, \bibinfo {author} {\bibfnamefont {A.}~\bibnamefont {Oyamada}},
  \bibinfo {author} {\bibfnamefont {S.}~\bibnamefont {Maegawa}}, \bibinfo
  {author} {\bibfnamefont {M.}~\bibnamefont {Tamura}},\ and\ \bibinfo {author}
  {\bibfnamefont {R.}~\bibnamefont {Kato}},\ }\href
  {https://journals.aps.org/prb/abstract/10.1103/PhysRevB.77.104413} {\bibfield
   {journal} {\bibinfo  {journal} {Phys. Rev. B}\ }\textbf {\bibinfo {volume}
  {77}},\ \bibinfo {pages} {104413} (\bibinfo {year} {2008})}\BibitemShut
  {NoStop}%
\bibitem [{\citenamefont {Tamura}\ and\ \citenamefont {Kato}(2009)}]{Tamura}%
  \BibitemOpen
  \bibfield  {author} {\bibinfo {author} {\bibfnamefont {M.}~\bibnamefont
  {Tamura}}\ and\ \bibinfo {author} {\bibfnamefont {R.}~\bibnamefont {Kato}},\
  }\href {http://dx.doi.org/10.1088/1468-6996/10/2/024304} {\bibfield
  {journal} {\bibinfo  {journal} {Sci. Technol. Adv. Mater.}\ }\textbf
  {\bibinfo {volume} {10}},\ \bibinfo {pages} {024304} (\bibinfo {year}
  {2009})}\BibitemShut {NoStop}%
\bibitem [{\citenamefont {Kanoda}\ and\ \citenamefont
  {Kato}(2011)}]{KanodaKato}%
  \BibitemOpen
  \bibfield  {author} {\bibinfo {author} {\bibfnamefont {K.}~\bibnamefont
  {Kanoda}}\ and\ \bibinfo {author} {\bibfnamefont {R.}~\bibnamefont {Kato}},\
  }\href
  {http://annualreviews.org/doi/abs/10.1146/annurev-conmatphys-062910-140521}
  {\bibfield  {journal} {\bibinfo  {journal} {Annu. Rev. Condens. Matter
  Phys.}\ }\textbf {\bibinfo {volume} {2}},\ \bibinfo {pages} {167} (\bibinfo
  {year} {2011})}\BibitemShut {NoStop}%
\bibitem [{\citenamefont {Kato}(2014)}]{KatoBulletin}%
  \BibitemOpen
  \bibfield  {author} {\bibinfo {author} {\bibfnamefont {R.}~\bibnamefont
  {Kato}},\ }\href {https://doi.org/10.1246/bcsj.20130290} {\bibfield
  {journal} {\bibinfo  {journal} {Bull. Chem. Soc. Jpn.}\ }\textbf {\bibinfo
  {volume} {87}},\ \bibinfo {pages} {355} (\bibinfo {year} {2014})}\BibitemShut
  {NoStop}%
\bibitem [{\citenamefont {Ueda}\ \emph {et~al.}(2018)\citenamefont {Ueda},
  \citenamefont {Tsumuraya},\ and\ \citenamefont {Kato}}]{Ueda}%
  \BibitemOpen
  \bibfield  {author} {\bibinfo {author} {\bibfnamefont {K.}~\bibnamefont
  {Ueda}}, \bibinfo {author} {\bibfnamefont {T.}~\bibnamefont {Tsumuraya}},\
  and\ \bibinfo {author} {\bibfnamefont {R.}~\bibnamefont {Kato}},\ }\href
  {http://dx.doi.org/10.3390/cryst8030138} {\bibfield  {journal} {\bibinfo
  {journal} {Crystals}\ }\textbf {\bibinfo {volume} {8}},\ \bibinfo {pages}
  {138} (\bibinfo {year} {2018})}\BibitemShut {NoStop}%
\bibitem [{\citenamefont {Nakamura}\ \emph {et~al.}(2012)\citenamefont
  {Nakamura}, \citenamefont {Yoshimoto},\ and\ \citenamefont
  {Imada}}]{NakamuraPRB}%
  \BibitemOpen
  \bibfield  {author} {\bibinfo {author} {\bibfnamefont {K.}~\bibnamefont
  {Nakamura}}, \bibinfo {author} {\bibfnamefont {Y.}~\bibnamefont
  {Yoshimoto}},\ and\ \bibinfo {author} {\bibfnamefont {M.}~\bibnamefont
  {Imada}},\ }\href {https://doi.org/ARTN 205117 10.1103/PhysRevB.86.205117}
  {\bibfield  {journal} {\bibinfo  {journal} {Physical Review B}\ }\textbf
  {\bibinfo {volume} {86}},\ \bibinfo {pages} {205117} (\bibinfo {year}
  {2012})}\BibitemShut {NoStop}%
\bibitem [{\citenamefont {Scriven}\ and\ \citenamefont
  {Powell}(2012)}]{ScrivenPRL}%
  \BibitemOpen
  \bibfield  {author} {\bibinfo {author} {\bibfnamefont {E.~P.}\ \bibnamefont
  {Scriven}}\ and\ \bibinfo {author} {\bibfnamefont {B.~J.}\ \bibnamefont
  {Powell}},\ }\href {https://doi.org/10.1103/PhysRevLett.109.097206}
  {\bibfield  {journal} {\bibinfo  {journal} {Phys Rev Lett}\ }\textbf
  {\bibinfo {volume} {109}},\ \bibinfo {pages} {097206} (\bibinfo {year}
  {2012})}\BibitemShut {NoStop}%
\bibitem [{\citenamefont {Tsumuraya}\ \emph {et~al.}(2013)\citenamefont
  {Tsumuraya}, \citenamefont {Seo}, \citenamefont {Tsuchiizu}, \citenamefont
  {Kato},\ and\ \citenamefont {Miyazaki}}]{Tsumuraya}%
  \BibitemOpen
  \bibfield  {author} {\bibinfo {author} {\bibfnamefont {T.}~\bibnamefont
  {Tsumuraya}}, \bibinfo {author} {\bibfnamefont {H.}~\bibnamefont {Seo}},
  \bibinfo {author} {\bibfnamefont {M.}~\bibnamefont {Tsuchiizu}}, \bibinfo
  {author} {\bibfnamefont {R.}~\bibnamefont {Kato}},\ and\ \bibinfo {author}
  {\bibfnamefont {T.}~\bibnamefont {Miyazaki}},\ }\href
  {http://journals.jps.jp/doi/abs/10.7566/JPSJ.82.033709} {\bibfield  {journal}
  {\bibinfo  {journal} {J. Phys. Soc. Jpn.}\ }\textbf {\bibinfo {volume}
  {82}},\ \bibinfo {pages} {033709} (\bibinfo {year} {2013})}\BibitemShut
  {NoStop}%
\bibitem [{\citenamefont {Yoshimi}\ \emph {et~al.}(2021)\citenamefont
  {Yoshimi}, \citenamefont {Tsumuraya},\ and\ \citenamefont
  {Misawa}}]{YoshimiPRR}%
  \BibitemOpen
  \bibfield  {author} {\bibinfo {author} {\bibfnamefont {K.}~\bibnamefont
  {Yoshimi}}, \bibinfo {author} {\bibfnamefont {T.}~\bibnamefont {Tsumuraya}},\
  and\ \bibinfo {author} {\bibfnamefont {T.}~\bibnamefont {Misawa}},\ }\href
  {https://doi.org/ARTN 043224 10.1103/PhysRevResearch.3.043224} {\bibfield
  {journal} {\bibinfo  {journal} {Physical Review Research}\ }\textbf {\bibinfo
  {volume} {3}},\ \bibinfo {pages} {043224} (\bibinfo {year}
  {2021})}\BibitemShut {NoStop}%
\bibitem [{\citenamefont {Ido}\ \emph {et~al.}(2022)\citenamefont {Ido},
  \citenamefont {Yoshimi}, \citenamefont {Misawa},\ and\ \citenamefont
  {Imada}}]{IdoQM}%
  \BibitemOpen
  \bibfield  {author} {\bibinfo {author} {\bibfnamefont {K.}~\bibnamefont
  {Ido}}, \bibinfo {author} {\bibfnamefont {K.}~\bibnamefont {Yoshimi}},
  \bibinfo {author} {\bibfnamefont {T.}~\bibnamefont {Misawa}},\ and\ \bibinfo
  {author} {\bibfnamefont {M.}~\bibnamefont {Imada}},\ }\href
  {https://doi.org/10.1038/s41535-022-00452-8} {\bibfield  {journal} {\bibinfo
  {journal} {npj Quantum Materials}\ }\textbf {\bibinfo {volume} {7}},\
  \bibinfo {pages} {48} (\bibinfo {year} {2022})}\BibitemShut {NoStop}%
\bibitem [{\citenamefont {Itou}\ \emph {et~al.}(2010)\citenamefont {Itou},
  \citenamefont {Oyamada}, \citenamefont {Maegawa},\ and\ \citenamefont
  {Kato}}]{ItouNatPhys}%
  \BibitemOpen
  \bibfield  {author} {\bibinfo {author} {\bibfnamefont {T.}~\bibnamefont
  {Itou}}, \bibinfo {author} {\bibfnamefont {A.}~\bibnamefont {Oyamada}},
  \bibinfo {author} {\bibfnamefont {S.}~\bibnamefont {Maegawa}},\ and\ \bibinfo
  {author} {\bibfnamefont {R.}~\bibnamefont {Kato}},\ }\href
  {http://dx.doi.org/10.1038/nphys1715
  https://www.nature.com/articles/nphys1715.pdf} {\bibfield  {journal}
  {\bibinfo  {journal} {Nat. Phys.}\ }\textbf {\bibinfo {volume} {6}},\
  \bibinfo {pages} {673} (\bibinfo {year} {2010})}\BibitemShut {NoStop}%
\bibitem [{\citenamefont {Yamashita}\ \emph {et~al.}(2010)\citenamefont
  {Yamashita}, \citenamefont {Nakata}, \citenamefont {Senshu}, \citenamefont
  {Nagata}, \citenamefont {Yamamoto}, \citenamefont {Kato}, \citenamefont
  {Shibauchi},\ and\ \citenamefont {Matsuda}}]{MinoruScience}%
  \BibitemOpen
  \bibfield  {author} {\bibinfo {author} {\bibfnamefont {M.}~\bibnamefont
  {Yamashita}}, \bibinfo {author} {\bibfnamefont {N.}~\bibnamefont {Nakata}},
  \bibinfo {author} {\bibfnamefont {Y.}~\bibnamefont {Senshu}}, \bibinfo
  {author} {\bibfnamefont {M.}~\bibnamefont {Nagata}}, \bibinfo {author}
  {\bibfnamefont {H.~M.}\ \bibnamefont {Yamamoto}}, \bibinfo {author}
  {\bibfnamefont {R.}~\bibnamefont {Kato}}, \bibinfo {author} {\bibfnamefont
  {T.}~\bibnamefont {Shibauchi}},\ and\ \bibinfo {author} {\bibfnamefont
  {Y.}~\bibnamefont {Matsuda}},\ }\href
  {http://dx.doi.org/10.1126/science.1188200
  https://science.sciencemag.org/content/sci/328/5983/1246.full.pdf} {\bibfield
   {journal} {\bibinfo  {journal} {Science}\ }\textbf {\bibinfo {volume}
  {328}},\ \bibinfo {pages} {1246} (\bibinfo {year} {2010})}\BibitemShut
  {NoStop}%
\bibitem [{\citenamefont {Yamashita}\ \emph {et~al.}(2011)\citenamefont
  {Yamashita}, \citenamefont {Yamamoto}, \citenamefont {Nakazawa},
  \citenamefont {Tamura},\ and\ \citenamefont {Kato}}]{SatoshiNatComm}%
  \BibitemOpen
  \bibfield  {author} {\bibinfo {author} {\bibfnamefont {S.}~\bibnamefont
  {Yamashita}}, \bibinfo {author} {\bibfnamefont {T.}~\bibnamefont {Yamamoto}},
  \bibinfo {author} {\bibfnamefont {Y.}~\bibnamefont {Nakazawa}}, \bibinfo
  {author} {\bibfnamefont {M.}~\bibnamefont {Tamura}},\ and\ \bibinfo {author}
  {\bibfnamefont {R.}~\bibnamefont {Kato}},\ }\href
  {http://dx.doi.org/10.1038/ncomms1274
  https://www.ncbi.nlm.nih.gov/pmc/articles/PMC3104558/pdf/ncomms1274.pdf}
  {\bibfield  {journal} {\bibinfo  {journal} {Nat. Commun.}\ }\textbf {\bibinfo
  {volume} {2}},\ \bibinfo {pages} {275} (\bibinfo {year} {2011})}\BibitemShut
  {NoStop}%
\bibitem [{\citenamefont {Watanabe}\ \emph {et~al.}(2012)\citenamefont
  {Watanabe}, \citenamefont {Yamashita}, \citenamefont {Tonegawa},
  \citenamefont {Oshima}, \citenamefont {Yamamoto}, \citenamefont {Kato},
  \citenamefont {Sheikin}, \citenamefont {Behnia}, \citenamefont {Terashima},
  \citenamefont {Uji}, \citenamefont {Shibauchi},\ and\ \citenamefont
  {Matsuda}}]{WatanabeNatComm}%
  \BibitemOpen
  \bibfield  {author} {\bibinfo {author} {\bibfnamefont {D.}~\bibnamefont
  {Watanabe}}, \bibinfo {author} {\bibfnamefont {M.}~\bibnamefont {Yamashita}},
  \bibinfo {author} {\bibfnamefont {S.}~\bibnamefont {Tonegawa}}, \bibinfo
  {author} {\bibfnamefont {Y.}~\bibnamefont {Oshima}}, \bibinfo {author}
  {\bibfnamefont {H.~M.}\ \bibnamefont {Yamamoto}}, \bibinfo {author}
  {\bibfnamefont {R.}~\bibnamefont {Kato}}, \bibinfo {author} {\bibfnamefont
  {I.}~\bibnamefont {Sheikin}}, \bibinfo {author} {\bibfnamefont
  {K.}~\bibnamefont {Behnia}}, \bibinfo {author} {\bibfnamefont
  {T.}~\bibnamefont {Terashima}}, \bibinfo {author} {\bibfnamefont
  {S.}~\bibnamefont {Uji}}, \bibinfo {author} {\bibfnamefont {T.}~\bibnamefont
  {Shibauchi}},\ and\ \bibinfo {author} {\bibfnamefont {Y.}~\bibnamefont
  {Matsuda}},\ }\href {<Go to ISI>://WOS:000309338100057
  http://www.nature.com/ncomms/journal/v3/n9/pdf/ncomms2082.pdf} {\bibfield
  {journal} {\bibinfo  {journal} {Nat. Commun.}\ }\textbf {\bibinfo {volume}
  {3}},\ \bibinfo {pages} {1090} (\bibinfo {year} {2012})}\BibitemShut
  {NoStop}%
\bibitem [{\citenamefont {Poirier}\ \emph {et~al.}(2014)\citenamefont
  {Poirier}, \citenamefont {Proulx},\ and\ \citenamefont {Kato}}]{Poirier}%
  \BibitemOpen
  \bibfield  {author} {\bibinfo {author} {\bibfnamefont {M.}~\bibnamefont
  {Poirier}}, \bibinfo {author} {\bibfnamefont {M.~O.}\ \bibnamefont
  {Proulx}},\ and\ \bibinfo {author} {\bibfnamefont {R.}~\bibnamefont {Kato}},\
  }\href {https://journals.aps.org/prb/abstract/10.1103/PhysRevB.90.045147}
  {\bibfield  {journal} {\bibinfo  {journal} {Phys. Rev. B}\ }\textbf {\bibinfo
  {volume} {90}},\ \bibinfo {pages} {045147} (\bibinfo {year}
  {2014})}\BibitemShut {NoStop}%
\bibitem [{\citenamefont {Yamamoto}\ \emph {et~al.}(2017)\citenamefont
  {Yamamoto}, \citenamefont {Fujimoto}, \citenamefont {Naito}, \citenamefont
  {Nakazawa}, \citenamefont {Tamura}, \citenamefont {Yakushi}, \citenamefont
  {Ikemoto}, \citenamefont {Moriwaki},\ and\ \citenamefont {Kato}}]{Yamataka}%
  \BibitemOpen
  \bibfield  {author} {\bibinfo {author} {\bibfnamefont {T.}~\bibnamefont
  {Yamamoto}}, \bibinfo {author} {\bibfnamefont {T.}~\bibnamefont {Fujimoto}},
  \bibinfo {author} {\bibfnamefont {T.}~\bibnamefont {Naito}}, \bibinfo
  {author} {\bibfnamefont {Y.}~\bibnamefont {Nakazawa}}, \bibinfo {author}
  {\bibfnamefont {M.}~\bibnamefont {Tamura}}, \bibinfo {author} {\bibfnamefont
  {K.}~\bibnamefont {Yakushi}}, \bibinfo {author} {\bibfnamefont
  {Y.}~\bibnamefont {Ikemoto}}, \bibinfo {author} {\bibfnamefont
  {T.}~\bibnamefont {Moriwaki}},\ and\ \bibinfo {author} {\bibfnamefont
  {R.}~\bibnamefont {Kato}},\ }\href
  {http://dx.doi.org/10.1038/s41598-017-13118-4} {\bibfield  {journal}
  {\bibinfo  {journal} {Scientific Reports}\ }\textbf {\bibinfo {volume} {7}},\
  \bibinfo {pages} {12930} (\bibinfo {year} {2017})}\BibitemShut {NoStop}%
\bibitem [{\citenamefont {Ni}\ \emph {et~al.}(2019)\citenamefont {Ni},
  \citenamefont {Pan}, \citenamefont {Song}, \citenamefont {Huang},
  \citenamefont {Zeng}, \citenamefont {Yu}, \citenamefont {Cheng},
  \citenamefont {Wang}, \citenamefont {Dai}, \citenamefont {Kato},\ and\
  \citenamefont {Li}}]{Ni}%
  \BibitemOpen
  \bibfield  {author} {\bibinfo {author} {\bibfnamefont {J.~M.}\ \bibnamefont
  {Ni}}, \bibinfo {author} {\bibfnamefont {B.~L.}\ \bibnamefont {Pan}},
  \bibinfo {author} {\bibfnamefont {B.~Q.}\ \bibnamefont {Song}}, \bibinfo
  {author} {\bibfnamefont {Y.~Y.}\ \bibnamefont {Huang}}, \bibinfo {author}
  {\bibfnamefont {J.~Y.}\ \bibnamefont {Zeng}}, \bibinfo {author}
  {\bibfnamefont {Y.~J.}\ \bibnamefont {Yu}}, \bibinfo {author} {\bibfnamefont
  {E.~J.}\ \bibnamefont {Cheng}}, \bibinfo {author} {\bibfnamefont {L.~S.}\
  \bibnamefont {Wang}}, \bibinfo {author} {\bibfnamefont {D.~Z.}\ \bibnamefont
  {Dai}}, \bibinfo {author} {\bibfnamefont {R.}~\bibnamefont {Kato}},\ and\
  \bibinfo {author} {\bibfnamefont {S.~Y.}\ \bibnamefont {Li}},\ }\href {<Go to
  ISI>://WOS:000501809100006
  https://journals.aps.org/prl/pdf/10.1103/PhysRevLett.123.247204} {\bibfield
  {journal} {\bibinfo  {journal} {Phys. Rev. Lett.}\ }\textbf {\bibinfo
  {volume} {123}},\ \bibinfo {pages} {247204} (\bibinfo {year}
  {2019})}\BibitemShut {NoStop}%
\bibitem [{\citenamefont {Bourgeois-Hope}\ \emph {et~al.}(2019)\citenamefont
  {Bourgeois-Hope}, \citenamefont {Laliberte}, \citenamefont {Lefrancois},
  \citenamefont {Grissonnanche}, \citenamefont {de~Cotret}, \citenamefont
  {Gordon}, \citenamefont {Kitou}, \citenamefont {Sawa}, \citenamefont {Cui},
  \citenamefont {Kato}, \citenamefont {Taillefer},\ and\ \citenamefont
  {Doiron-Leyraud}}]{BourgeoisHope}%
  \BibitemOpen
  \bibfield  {author} {\bibinfo {author} {\bibfnamefont {P.}~\bibnamefont
  {Bourgeois-Hope}}, \bibinfo {author} {\bibfnamefont {F.}~\bibnamefont
  {Laliberte}}, \bibinfo {author} {\bibfnamefont {E.}~\bibnamefont
  {Lefrancois}}, \bibinfo {author} {\bibfnamefont {G.}~\bibnamefont
  {Grissonnanche}}, \bibinfo {author} {\bibfnamefont {S.~R.}\ \bibnamefont
  {de~Cotret}}, \bibinfo {author} {\bibfnamefont {R.}~\bibnamefont {Gordon}},
  \bibinfo {author} {\bibfnamefont {S.}~\bibnamefont {Kitou}}, \bibinfo
  {author} {\bibfnamefont {H.}~\bibnamefont {Sawa}}, \bibinfo {author}
  {\bibfnamefont {H.}~\bibnamefont {Cui}}, \bibinfo {author} {\bibfnamefont
  {R.}~\bibnamefont {Kato}}, \bibinfo {author} {\bibfnamefont {L.}~\bibnamefont
  {Taillefer}},\ and\ \bibinfo {author} {\bibfnamefont {N.}~\bibnamefont
  {Doiron-Leyraud}},\ }\href {<Go to ISI>://WOS:000501808700001
  https://journals.aps.org/prx/pdf/10.1103/PhysRevX.9.041051} {\bibfield
  {journal} {\bibinfo  {journal} {Phys. Rev. X}\ }\textbf {\bibinfo {volume}
  {9}},\ \bibinfo {pages} {041051} (\bibinfo {year} {2019})}\BibitemShut
  {NoStop}%
\bibitem [{\citenamefont {Yamashita}(2019)}]{MinoruJPSJ}%
  \BibitemOpen
  \bibfield  {author} {\bibinfo {author} {\bibfnamefont {M.}~\bibnamefont
  {Yamashita}},\ }\href {<Go to ISI>://WOS:000477716800011} {\bibfield
  {journal} {\bibinfo  {journal} {J. Phys. Soc. Jpn.}\ }\textbf {\bibinfo
  {volume} {88}},\ \bibinfo {pages} {083702} (\bibinfo {year}
  {2019})}\BibitemShut {NoStop}%
\bibitem [{\citenamefont {Yamashita}\ \emph {et~al.}(2020)\citenamefont
  {Yamashita}, \citenamefont {Sato}, \citenamefont {Tominaga}, \citenamefont
  {Kasahara}, \citenamefont {Kasahara}, \citenamefont {Cui}, \citenamefont
  {Kato}, \citenamefont {Shibauchi},\ and\ \citenamefont
  {Matsuda}}]{MinoruPRB}%
  \BibitemOpen
  \bibfield  {author} {\bibinfo {author} {\bibfnamefont {M.}~\bibnamefont
  {Yamashita}}, \bibinfo {author} {\bibfnamefont {Y.}~\bibnamefont {Sato}},
  \bibinfo {author} {\bibfnamefont {T.}~\bibnamefont {Tominaga}}, \bibinfo
  {author} {\bibfnamefont {Y.}~\bibnamefont {Kasahara}}, \bibinfo {author}
  {\bibfnamefont {S.}~\bibnamefont {Kasahara}}, \bibinfo {author}
  {\bibfnamefont {H.}~\bibnamefont {Cui}}, \bibinfo {author} {\bibfnamefont
  {R.}~\bibnamefont {Kato}}, \bibinfo {author} {\bibfnamefont {T.}~\bibnamefont
  {Shibauchi}},\ and\ \bibinfo {author} {\bibfnamefont {Y.}~\bibnamefont
  {Matsuda}},\ }\href {<Go to ISI>://WOS:000527119600001
  https://journals.aps.org/prb/pdf/10.1103/PhysRevB.101.140407} {\bibfield
  {journal} {\bibinfo  {journal} {Phys. Rev. B}\ }\textbf {\bibinfo {volume}
  {101}},\ \bibinfo {pages} {140407(R)} (\bibinfo {year} {2020})}\BibitemShut
  {NoStop}%
\bibitem [{\citenamefont {Nomoto}\ \emph {et~al.}(2022)\citenamefont {Nomoto},
  \citenamefont {Yamashita}, \citenamefont {Akutsu}, \citenamefont {Nakazawa},\
  and\ \citenamefont {Kato}}]{NomotoPRB}%
  \BibitemOpen
  \bibfield  {author} {\bibinfo {author} {\bibfnamefont {T.}~\bibnamefont
  {Nomoto}}, \bibinfo {author} {\bibfnamefont {S.}~\bibnamefont {Yamashita}},
  \bibinfo {author} {\bibfnamefont {H.}~\bibnamefont {Akutsu}}, \bibinfo
  {author} {\bibfnamefont {Y.}~\bibnamefont {Nakazawa}},\ and\ \bibinfo
  {author} {\bibfnamefont {R.}~\bibnamefont {Kato}},\ }\href
  {https://journals.aps.org/prb/pdf/10.1103/PhysRevB.105.245133} {\bibfield
  {journal} {\bibinfo  {journal} {Phys. Rev. B}\ }\textbf {\bibinfo {volume}
  {105}},\ \bibinfo {pages} {245133} (\bibinfo {year} {2022})}\BibitemShut
  {NoStop}%
\bibitem [{\citenamefont {Ueda}\ \emph {et~al.}(2024)\citenamefont {Ueda},
  \citenamefont {Fujiyama},\ and\ \citenamefont {Kato}}]{UedaPRB}%
  \BibitemOpen
  \bibfield  {author} {\bibinfo {author} {\bibfnamefont {K.}~\bibnamefont
  {Ueda}}, \bibinfo {author} {\bibfnamefont {S.}~\bibnamefont {Fujiyama}},\
  and\ \bibinfo {author} {\bibfnamefont {R.}~\bibnamefont {Kato}},\ }\href
  {https://doi.org/10.1103/PhysRevB.109.L140401} {\bibfield  {journal}
  {\bibinfo  {journal} {Phys. Rev. B}\ }\textbf {\bibinfo {volume} {109}},\
  \bibinfo {pages} {L140401} (\bibinfo {year} {2024})}\BibitemShut {NoStop}%
\bibitem [{\citenamefont {Motrunich}(2005)}]{Motrunich}%
  \BibitemOpen
  \bibfield  {author} {\bibinfo {author} {\bibfnamefont {O.~I.}\ \bibnamefont
  {Motrunich}},\ }\href {http://dx.doi.org/10.1103/PhysRevB.72.045105}
  {\bibfield  {journal} {\bibinfo  {journal} {Phys. Rev. B}\ }\textbf {\bibinfo
  {volume} {72}},\ \bibinfo {pages} {045105} (\bibinfo {year}
  {2005})}\BibitemShut {NoStop}%
\bibitem [{\citenamefont {Yamashita}\ \emph {et~al.}(2022)\citenamefont
  {Yamashita}, \citenamefont {Sato}, \citenamefont {Kasahara}, \citenamefont
  {Kasahara}, \citenamefont {Shibauchi},\ and\ \citenamefont
  {Matsuda}}]{MinoruSciRep}%
  \BibitemOpen
  \bibfield  {author} {\bibinfo {author} {\bibfnamefont {M.}~\bibnamefont
  {Yamashita}}, \bibinfo {author} {\bibfnamefont {Y.}~\bibnamefont {Sato}},
  \bibinfo {author} {\bibfnamefont {Y.}~\bibnamefont {Kasahara}}, \bibinfo
  {author} {\bibfnamefont {S.}~\bibnamefont {Kasahara}}, \bibinfo {author}
  {\bibfnamefont {T.}~\bibnamefont {Shibauchi}},\ and\ \bibinfo {author}
  {\bibfnamefont {Y.}~\bibnamefont {Matsuda}},\ }\href
  {https://doi.org/10.1038/s41598-022-13155-8} {\bibfield  {journal} {\bibinfo
  {journal} {Sci. Rep.}\ }\textbf {\bibinfo {volume} {12}},\ \bibinfo {pages}
  {9187} (\bibinfo {year} {2022})}\BibitemShut {NoStop}%
\bibitem [{\citenamefont {Kato}\ \emph {et~al.}(2022)\citenamefont {Kato},
  \citenamefont {Uebe}, \citenamefont {Fujiyama},\ and\ \citenamefont
  {Cui}}]{NoCoolingDepPaper}%
  \BibitemOpen
  \bibfield  {author} {\bibinfo {author} {\bibfnamefont {R.}~\bibnamefont
  {Kato}}, \bibinfo {author} {\bibfnamefont {M.}~\bibnamefont {Uebe}}, \bibinfo
  {author} {\bibfnamefont {S.}~\bibnamefont {Fujiyama}},\ and\ \bibinfo
  {author} {\bibfnamefont {H.~B.}\ \bibnamefont {Cui}},\ }\href
  {https://doi.org/10.3390/cryst12010102} {\bibfield  {journal} {\bibinfo
  {journal} {Crystals}\ }\textbf {\bibinfo {volume} {12}},\ \bibinfo {pages}
  {102} (\bibinfo {year} {2022})}\BibitemShut {NoStop}%
\bibitem [{\citenamefont {Watanabe}\ \emph {et~al.}(2014)\citenamefont
  {Watanabe}, \citenamefont {Kawamura}, \citenamefont {Nakano},\ and\
  \citenamefont {Sakai}}]{randomSingletJPSJ}%
  \BibitemOpen
  \bibfield  {author} {\bibinfo {author} {\bibfnamefont {K.}~\bibnamefont
  {Watanabe}}, \bibinfo {author} {\bibfnamefont {H.}~\bibnamefont {Kawamura}},
  \bibinfo {author} {\bibfnamefont {H.}~\bibnamefont {Nakano}},\ and\ \bibinfo
  {author} {\bibfnamefont {T.}~\bibnamefont {Sakai}},\ }\href
  {https://doi.org/10.7566/Jpsj.83.034714} {\bibfield  {journal} {\bibinfo
  {journal} {J. Phys. Soc. Jpn.}\ }\textbf {\bibinfo {volume} {83}},\ \bibinfo
  {pages} {034714} (\bibinfo {year} {2014})}\BibitemShut {NoStop}%
\bibitem [{\citenamefont {Shimokawa}\ \emph {et~al.}(2015)\citenamefont
  {Shimokawa}, \citenamefont {Watanabe},\ and\ \citenamefont
  {Kawamura}}]{randomSingletPRB}%
  \BibitemOpen
  \bibfield  {author} {\bibinfo {author} {\bibfnamefont {T.}~\bibnamefont
  {Shimokawa}}, \bibinfo {author} {\bibfnamefont {K.}~\bibnamefont
  {Watanabe}},\ and\ \bibinfo {author} {\bibfnamefont {H.}~\bibnamefont
  {Kawamura}},\ }\href {https://doi.org/10.1103/PhysRevB.92.134407} {\bibfield
  {journal} {\bibinfo  {journal} {Phys. Rev. B}\ }\textbf {\bibinfo {volume}
  {92}},\ \bibinfo {pages} {134407} (\bibinfo {year} {2015})}\BibitemShut
  {NoStop}%
\bibitem [{\citenamefont {Abdel-Jawad}\ \emph {et~al.}(2013)\citenamefont
  {Abdel-Jawad}, \citenamefont {Tajima}, \citenamefont {Kato},\ and\
  \citenamefont {Terasaki}}]{MajedPRB}%
  \BibitemOpen
  \bibfield  {author} {\bibinfo {author} {\bibfnamefont {M.}~\bibnamefont
  {Abdel-Jawad}}, \bibinfo {author} {\bibfnamefont {N.}~\bibnamefont {Tajima}},
  \bibinfo {author} {\bibfnamefont {R.}~\bibnamefont {Kato}},\ and\ \bibinfo
  {author} {\bibfnamefont {I.}~\bibnamefont {Terasaki}},\ }\href
  {https://doi.org/10.1103/PhysRevB.88.075139} {\bibfield  {journal} {\bibinfo
  {journal} {Phys. Rev. B}\ }\textbf {\bibinfo {volume} {88}},\ \bibinfo
  {pages} {075139} (\bibinfo {year} {2013})}\BibitemShut {NoStop}%
\bibitem [{SI()}]{SI}%
  \BibitemOpen
  \href@noop {} {}\bibinfo {note} {URL will be inserted by
  publisher}\BibitemShut {NoStop}%
\bibitem [{\citenamefont {Anderson}(1954)}]{AndersonJPSJ}%
  \BibitemOpen
  \bibfield  {author} {\bibinfo {author} {\bibfnamefont {P.~W.}\ \bibnamefont
  {Anderson}},\ }\href {http://journals.jps.jp/doi/abs/10.1143/JPSJ.9.316}
  {\bibfield  {journal} {\bibinfo  {journal} {J. Phys. Soc. Jpn.}\ }\textbf
  {\bibinfo {volume} {9}},\ \bibinfo {pages} {316} (\bibinfo {year}
  {1954})}\BibitemShut {NoStop}%
\bibitem [{\citenamefont {Bencini}\ and\ \citenamefont
  {Gatteschi}(1990)}]{BenciniGatteschi}%
  \BibitemOpen
  \bibfield  {author} {\bibinfo {author} {\bibfnamefont {A.}~\bibnamefont
  {Bencini}}\ and\ \bibinfo {author} {\bibfnamefont {D.}~\bibnamefont
  {Gatteschi}},\ }\href@noop {} {\emph {\bibinfo {title} {Electron paramagnetic
  resonance of exchange coupled systems}}}\ (\bibinfo  {publisher} {Springer},\
  \bibinfo {address} {Berlin},\ \bibinfo {year} {1990})\BibitemShut {NoStop}%
\bibitem [{\citenamefont {Okuda}\ and\ \citenamefont {Date}(1970)}]{OkudaDate}%
  \BibitemOpen
  \bibfield  {author} {\bibinfo {author} {\bibfnamefont {T.}~\bibnamefont
  {Okuda}}\ and\ \bibinfo {author} {\bibfnamefont {M.}~\bibnamefont {Date}},\
  }\href {https://doi.org/Doi 10.1143/Jpsj.28.308} {\bibfield  {journal}
  {\bibinfo  {journal} {J. Phys. Soc. Jpn.}\ }\textbf {\bibinfo {volume}
  {28}},\ \bibinfo {pages} {308} (\bibinfo {year} {1970})}\BibitemShut
  {NoStop}%
\bibitem [{\citenamefont {Fayzullin}\ \emph {et~al.}(2013)\citenamefont
  {Fayzullin}, \citenamefont {Eremina}, \citenamefont {Eremin}, \citenamefont
  {Dittl}, \citenamefont {van Well}, \citenamefont {Ritter}, \citenamefont
  {Assmus}, \citenamefont {Deisenhofer}, \citenamefont {von Nidda},\ and\
  \citenamefont {Loidl}}]{Fayzullin}%
  \BibitemOpen
  \bibfield  {author} {\bibinfo {author} {\bibfnamefont {M.~A.}\ \bibnamefont
  {Fayzullin}}, \bibinfo {author} {\bibfnamefont {R.~M.}\ \bibnamefont
  {Eremina}}, \bibinfo {author} {\bibfnamefont {M.~V.}\ \bibnamefont {Eremin}},
  \bibinfo {author} {\bibfnamefont {A.}~\bibnamefont {Dittl}}, \bibinfo
  {author} {\bibfnamefont {N.}~\bibnamefont {van Well}}, \bibinfo {author}
  {\bibfnamefont {F.}~\bibnamefont {Ritter}}, \bibinfo {author} {\bibfnamefont
  {W.}~\bibnamefont {Assmus}}, \bibinfo {author} {\bibfnamefont
  {J.}~\bibnamefont {Deisenhofer}}, \bibinfo {author} {\bibfnamefont
  {H.-A.~K.}\ \bibnamefont {von Nidda}},\ and\ \bibinfo {author} {\bibfnamefont
  {A.}~\bibnamefont {Loidl}},\ }\href
  {http://dx.doi.org/10.1103/PhysRevB.88.174421} {\bibfield  {journal}
  {\bibinfo  {journal} {Phys. Rev. B}\ }\textbf {\bibinfo {volume} {88}},\
  \bibinfo {pages} {174421} (\bibinfo {year} {2013})}\BibitemShut {NoStop}%
\bibitem [{\citenamefont {Hennessy}\ \emph {et~al.}(1973)\citenamefont
  {Hennessy}, \citenamefont {McElwee},\ and\ \citenamefont
  {Richards}}]{Hennessy}%
  \BibitemOpen
  \bibfield  {author} {\bibinfo {author} {\bibfnamefont {M.~J.}\ \bibnamefont
  {Hennessy}}, \bibinfo {author} {\bibfnamefont {C.~D.}\ \bibnamefont
  {McElwee}},\ and\ \bibinfo {author} {\bibfnamefont {P.~M.}\ \bibnamefont
  {Richards}},\ }\href {<Go to ISI>://WOS:A1973O668400007
  https://journals.aps.org/prb/pdf/10.1103/PhysRevB.7.930} {\bibfield
  {journal} {\bibinfo  {journal} {Phys. Rev. B}\ }\textbf {\bibinfo {volume}
  {7}},\ \bibinfo {pages} {930} (\bibinfo {year} {1973})}\BibitemShut {NoStop}%
\bibitem [{\citenamefont {Cheung}\ \emph {et~al.}(1978)\citenamefont {Cheung},
  \citenamefont {Soos}, \citenamefont {Dietz},\ and\ \citenamefont
  {Merritt}}]{Cheung}%
  \BibitemOpen
  \bibfield  {author} {\bibinfo {author} {\bibfnamefont {T.~T.~P.}\
  \bibnamefont {Cheung}}, \bibinfo {author} {\bibfnamefont {Z.~G.}\
  \bibnamefont {Soos}}, \bibinfo {author} {\bibfnamefont {R.~E.}\ \bibnamefont
  {Dietz}},\ and\ \bibinfo {author} {\bibfnamefont {F.~R.}\ \bibnamefont
  {Merritt}},\ }\href {<Go to ISI>://WOS:A1978ET38000031
  https://journals.aps.org/prb/pdf/10.1103/PhysRevB.17.1266} {\bibfield
  {journal} {\bibinfo  {journal} {Phys. Rev. B}\ }\textbf {\bibinfo {volume}
  {17}},\ \bibinfo {pages} {1266} (\bibinfo {year} {1978})}\BibitemShut
  {NoStop}%
\bibitem [{\citenamefont {Takahashi}\ \emph {et~al.}(1980)\citenamefont
  {Takahashi}, \citenamefont {Doi},\ and\ \citenamefont
  {Nagasawa}}]{Takahashi}%
  \BibitemOpen
  \bibfield  {author} {\bibinfo {author} {\bibfnamefont {T.}~\bibnamefont
  {Takahashi}}, \bibinfo {author} {\bibfnamefont {H.}~\bibnamefont {Doi}},\
  and\ \bibinfo {author} {\bibfnamefont {H.}~\bibnamefont {Nagasawa}},\ }\href
  {<Go to ISI>://WOS:A1980JM64000010} {\bibfield  {journal} {\bibinfo
  {journal} {J. Phys. Soc. Jpn.}\ }\textbf {\bibinfo {volume} {48}},\ \bibinfo
  {pages} {423} (\bibinfo {year} {1980})}\BibitemShut {NoStop}%
\bibitem [{\citenamefont {Tanaka}\ \emph {et~al.}(2006)\citenamefont {Tanaka},
  \citenamefont {Kuroda}, \citenamefont {Yamashita}, \citenamefont {Mitsumi},\
  and\ \citenamefont {Toriumi}}]{Tanaka}%
  \BibitemOpen
  \bibfield  {author} {\bibinfo {author} {\bibfnamefont {H.}~\bibnamefont
  {Tanaka}}, \bibinfo {author} {\bibfnamefont {S.~I.}\ \bibnamefont {Kuroda}},
  \bibinfo {author} {\bibfnamefont {T.}~\bibnamefont {Yamashita}}, \bibinfo
  {author} {\bibfnamefont {M.}~\bibnamefont {Mitsumi}},\ and\ \bibinfo {author}
  {\bibfnamefont {K.}~\bibnamefont {Toriumi}},\ }\href {<Go to
  ISI>://WOS:000238696900026} {\bibfield  {journal} {\bibinfo  {journal} {Phys.
  Rev. B}\ }\textbf {\bibinfo {volume} {73}},\ \bibinfo {pages} {245102}
  (\bibinfo {year} {2006})}\BibitemShut {NoStop}%
\bibitem [{\citenamefont {Frisch}\ \emph {et~al.}(2016)\citenamefont {Frisch},
  \citenamefont {Trucks}, \citenamefont {Schlegel}, \citenamefont {Scuseria},
  \citenamefont {Robb}, \citenamefont {Cheeseman}, \citenamefont {Scalmani},
  \citenamefont {Barone}, \citenamefont {Petersson}, \citenamefont {Nakatsuji},
  \citenamefont {Li}, \citenamefont {Caricato}, \citenamefont {Marenich},
  \citenamefont {Bloino}, \citenamefont {Janesko}, \citenamefont {Gomperts},
  \citenamefont {Mennucci}, \citenamefont {Hratchian}, \citenamefont {Ortiz},
  \citenamefont {Izmaylov}, \citenamefont {Sonnenberg}, \citenamefont
  {Williams}, \citenamefont {Ding}, \citenamefont {Lipparini}, \citenamefont
  {Egidi}, \citenamefont {Goings}, \citenamefont {Peng}, \citenamefont
  {Petrone}, \citenamefont {Henderson}, \citenamefont {Ranasinghe},
  \citenamefont {Zakrzewski}, \citenamefont {Gao}, \citenamefont {Rega},
  \citenamefont {Zheng}, \citenamefont {Liang}, \citenamefont {Hada},
  \citenamefont {Ehara}, \citenamefont {Toyota}, \citenamefont {Fukuda},
  \citenamefont {Hasegawa}, \citenamefont {Ishida}, \citenamefont {Nakajima},
  \citenamefont {Honda}, \citenamefont {Kitao}, \citenamefont {Nakai},
  \citenamefont {Vreven}, \citenamefont {Throssell}, \citenamefont
  {Montgomery~Jr.}, \citenamefont {Peralta}, \citenamefont {Ogliaro},
  \citenamefont {Bearpark}, \citenamefont {Heyd}, \citenamefont {Brothers},
  \citenamefont {Kudin}, \citenamefont {Staroverov}, \citenamefont {Keith},
  \citenamefont {Kobayashi}, \citenamefont {Normand}, \citenamefont
  {Raghavachari}, \citenamefont {Rendell}, \citenamefont {Burant},
  \citenamefont {Iyengar}, \citenamefont {Tomasi}, \citenamefont {Cossi},
  \citenamefont {Millam}, \citenamefont {Klene}, \citenamefont {Adamo},
  \citenamefont {Cammi}, \citenamefont {Ochterski}, \citenamefont {Martin},
  \citenamefont {Morokuma}, \citenamefont {Farkas}, \citenamefont {Foresman},\
  and\ \citenamefont {Fox}}]{Gaussian}%
  \BibitemOpen
  \bibfield  {author} {\bibinfo {author} {\bibfnamefont {M.~J.}\ \bibnamefont
  {Frisch}}, \bibinfo {author} {\bibfnamefont {G.~W.}\ \bibnamefont {Trucks}},
  \bibinfo {author} {\bibfnamefont {H.~B.}\ \bibnamefont {Schlegel}}, \bibinfo
  {author} {\bibfnamefont {G.~E.}\ \bibnamefont {Scuseria}}, \bibinfo {author}
  {\bibfnamefont {M.~A.}\ \bibnamefont {Robb}}, \bibinfo {author}
  {\bibfnamefont {J.~R.}\ \bibnamefont {Cheeseman}}, \bibinfo {author}
  {\bibfnamefont {G.}~\bibnamefont {Scalmani}}, \bibinfo {author}
  {\bibfnamefont {V.}~\bibnamefont {Barone}}, \bibinfo {author} {\bibfnamefont
  {G.~A.}\ \bibnamefont {Petersson}}, \bibinfo {author} {\bibfnamefont
  {H.}~\bibnamefont {Nakatsuji}}, \bibinfo {author} {\bibfnamefont
  {X.}~\bibnamefont {Li}}, \bibinfo {author} {\bibfnamefont {M.}~\bibnamefont
  {Caricato}}, \bibinfo {author} {\bibfnamefont {A.~V.}\ \bibnamefont
  {Marenich}}, \bibinfo {author} {\bibfnamefont {J.}~\bibnamefont {Bloino}},
  \bibinfo {author} {\bibfnamefont {B.~G.}\ \bibnamefont {Janesko}}, \bibinfo
  {author} {\bibfnamefont {R.}~\bibnamefont {Gomperts}}, \bibinfo {author}
  {\bibfnamefont {B.}~\bibnamefont {Mennucci}}, \bibinfo {author}
  {\bibfnamefont {H.~P.}\ \bibnamefont {Hratchian}}, \bibinfo {author}
  {\bibfnamefont {J.~V.}\ \bibnamefont {Ortiz}}, \bibinfo {author}
  {\bibfnamefont {A.~F.}\ \bibnamefont {Izmaylov}}, \bibinfo {author}
  {\bibfnamefont {J.~L.}\ \bibnamefont {Sonnenberg}}, \bibinfo {author}
  {\bibnamefont {Williams}}, \bibinfo {author} {\bibfnamefont {F.}~\bibnamefont
  {Ding}}, \bibinfo {author} {\bibfnamefont {F.}~\bibnamefont {Lipparini}},
  \bibinfo {author} {\bibfnamefont {F.}~\bibnamefont {Egidi}}, \bibinfo
  {author} {\bibfnamefont {J.}~\bibnamefont {Goings}}, \bibinfo {author}
  {\bibfnamefont {B.}~\bibnamefont {Peng}}, \bibinfo {author} {\bibfnamefont
  {A.}~\bibnamefont {Petrone}}, \bibinfo {author} {\bibfnamefont
  {T.}~\bibnamefont {Henderson}}, \bibinfo {author} {\bibfnamefont
  {D.}~\bibnamefont {Ranasinghe}}, \bibinfo {author} {\bibfnamefont {V.~G.}\
  \bibnamefont {Zakrzewski}}, \bibinfo {author} {\bibfnamefont
  {J.}~\bibnamefont {Gao}}, \bibinfo {author} {\bibfnamefont {N.}~\bibnamefont
  {Rega}}, \bibinfo {author} {\bibfnamefont {G.}~\bibnamefont {Zheng}},
  \bibinfo {author} {\bibfnamefont {W.}~\bibnamefont {Liang}}, \bibinfo
  {author} {\bibfnamefont {M.}~\bibnamefont {Hada}}, \bibinfo {author}
  {\bibfnamefont {M.}~\bibnamefont {Ehara}}, \bibinfo {author} {\bibfnamefont
  {K.}~\bibnamefont {Toyota}}, \bibinfo {author} {\bibfnamefont
  {R.}~\bibnamefont {Fukuda}}, \bibinfo {author} {\bibfnamefont
  {J.}~\bibnamefont {Hasegawa}}, \bibinfo {author} {\bibfnamefont
  {M.}~\bibnamefont {Ishida}}, \bibinfo {author} {\bibfnamefont
  {T.}~\bibnamefont {Nakajima}}, \bibinfo {author} {\bibfnamefont
  {Y.}~\bibnamefont {Honda}}, \bibinfo {author} {\bibfnamefont
  {O.}~\bibnamefont {Kitao}}, \bibinfo {author} {\bibfnamefont
  {H.}~\bibnamefont {Nakai}}, \bibinfo {author} {\bibfnamefont
  {T.}~\bibnamefont {Vreven}}, \bibinfo {author} {\bibfnamefont
  {K.}~\bibnamefont {Throssell}}, \bibinfo {author} {\bibfnamefont {J.~A.}\
  \bibnamefont {Montgomery~Jr.}}, \bibinfo {author} {\bibfnamefont {J.~E.}\
  \bibnamefont {Peralta}}, \bibinfo {author} {\bibfnamefont {F.}~\bibnamefont
  {Ogliaro}}, \bibinfo {author} {\bibfnamefont {M.~J.}\ \bibnamefont
  {Bearpark}}, \bibinfo {author} {\bibfnamefont {J.~J.}\ \bibnamefont {Heyd}},
  \bibinfo {author} {\bibfnamefont {E.~N.}\ \bibnamefont {Brothers}}, \bibinfo
  {author} {\bibfnamefont {K.~N.}\ \bibnamefont {Kudin}}, \bibinfo {author}
  {\bibfnamefont {V.~N.}\ \bibnamefont {Staroverov}}, \bibinfo {author}
  {\bibfnamefont {T.~A.}\ \bibnamefont {Keith}}, \bibinfo {author}
  {\bibfnamefont {R.}~\bibnamefont {Kobayashi}}, \bibinfo {author}
  {\bibfnamefont {J.}~\bibnamefont {Normand}}, \bibinfo {author} {\bibfnamefont
  {K.}~\bibnamefont {Raghavachari}}, \bibinfo {author} {\bibfnamefont {A.~P.}\
  \bibnamefont {Rendell}}, \bibinfo {author} {\bibfnamefont {J.~C.}\
  \bibnamefont {Burant}}, \bibinfo {author} {\bibfnamefont {S.~S.}\
  \bibnamefont {Iyengar}}, \bibinfo {author} {\bibfnamefont {J.}~\bibnamefont
  {Tomasi}}, \bibinfo {author} {\bibfnamefont {M.}~\bibnamefont {Cossi}},
  \bibinfo {author} {\bibfnamefont {J.~M.}\ \bibnamefont {Millam}}, \bibinfo
  {author} {\bibfnamefont {M.}~\bibnamefont {Klene}}, \bibinfo {author}
  {\bibfnamefont {C.}~\bibnamefont {Adamo}}, \bibinfo {author} {\bibfnamefont
  {R.}~\bibnamefont {Cammi}}, \bibinfo {author} {\bibfnamefont {J.~W.}\
  \bibnamefont {Ochterski}}, \bibinfo {author} {\bibfnamefont {R.~L.}\
  \bibnamefont {Martin}}, \bibinfo {author} {\bibfnamefont {K.}~\bibnamefont
  {Morokuma}}, \bibinfo {author} {\bibfnamefont {O.}~\bibnamefont {Farkas}},
  \bibinfo {author} {\bibfnamefont {J.~B.}\ \bibnamefont {Foresman}},\ and\
  \bibinfo {author} {\bibfnamefont {D.~J.}\ \bibnamefont {Fox}},\ }\href@noop
  {} {\bibinfo {title} {Gaussian 16 rev. b.01}} (\bibinfo {year}
  {2016})\BibitemShut {NoStop}%
\bibitem [{\citenamefont {Pratt}\ \emph {et~al.}(2000)\citenamefont {Pratt},
  \citenamefont {Blundell}, \citenamefont {Jestadt}, \citenamefont {Lovett},
  \citenamefont {Macrae},\ and\ \citenamefont {Hayes}}]{PrattMRC}%
  \BibitemOpen
  \bibfield  {author} {\bibinfo {author} {\bibfnamefont {F.~L.}\ \bibnamefont
  {Pratt}}, \bibinfo {author} {\bibfnamefont {S.~J.}\ \bibnamefont {Blundell}},
  \bibinfo {author} {\bibfnamefont {T.}~\bibnamefont {Jestadt}}, \bibinfo
  {author} {\bibfnamefont {B.~W.}\ \bibnamefont {Lovett}}, \bibinfo {author}
  {\bibfnamefont {R.~M.}\ \bibnamefont {Macrae}},\ and\ \bibinfo {author}
  {\bibfnamefont {W.}~\bibnamefont {Hayes}},\ }\href {<Go to
  ISI>://WOS:000087836300006
  https://onlinelibrary.wiley.com/doi/pdfdirect/10.1002/1097-458X%28200006%2938%3A13%3C%3A%3AAID-MRC695%3E3.0.CO%3B2-W?download=true}
  {\bibfield  {journal} {\bibinfo  {journal} {Magn. Reson. Chem.}\ }\textbf
  {\bibinfo {volume} {38}},\ \bibinfo {pages} {S27} (\bibinfo {year}
  {2000})}\BibitemShut {NoStop}%
\bibitem [{\citenamefont {Momma}\ and\ \citenamefont {Izumi}(2011)}]{VESTA}%
  \BibitemOpen
  \bibfield  {author} {\bibinfo {author} {\bibfnamefont {K.}~\bibnamefont
  {Momma}}\ and\ \bibinfo {author} {\bibfnamefont {F.}~\bibnamefont {Izumi}},\
  }\href@noop {} {\bibfield  {journal} {\bibinfo  {journal} {J. Appl.
  Crystallogr.}\ }\textbf {\bibinfo {volume} {44}},\ \bibinfo {pages} {1272}
  (\bibinfo {year} {2011})}\BibitemShut {NoStop}%
\bibitem [{\citenamefont {Misawa}\ \emph {et~al.}(2020)\citenamefont {Misawa},
  \citenamefont {Yoshimi},\ and\ \citenamefont {Tsumuraya}}]{MisawaPRR}%
  \BibitemOpen
  \bibfield  {author} {\bibinfo {author} {\bibfnamefont {T.}~\bibnamefont
  {Misawa}}, \bibinfo {author} {\bibfnamefont {K.}~\bibnamefont {Yoshimi}},\
  and\ \bibinfo {author} {\bibfnamefont {T.}~\bibnamefont {Tsumuraya}},\ }\href
  {https://doi.org/10.1103/PhysRevResearch.2.032072} {\bibfield  {journal}
  {\bibinfo  {journal} {Phys. Rev. Res.}\ }\textbf {\bibinfo {volume} {2}},\
  \bibinfo {pages} {032072(R)} (\bibinfo {year} {2020})}\BibitemShut {NoStop}%
\bibitem [{\citenamefont {Perdew}\ \emph {et~al.}(1996)\citenamefont {Perdew},
  \citenamefont {Burke},\ and\ \citenamefont {Ernzerhof}}]{PerdewPRL}%
  \BibitemOpen
  \bibfield  {author} {\bibinfo {author} {\bibfnamefont {J.~P.}\ \bibnamefont
  {Perdew}}, \bibinfo {author} {\bibfnamefont {K.}~\bibnamefont {Burke}},\ and\
  \bibinfo {author} {\bibfnamefont {M.}~\bibnamefont {Ernzerhof}},\ }\href
  {https://doi.org/10.1103/PhysRevLett.77.3865} {\bibfield  {journal} {\bibinfo
   {journal} {Phys. Rev. Lett.}\ }\textbf {\bibinfo {volume} {77}},\ \bibinfo
  {pages} {3865} (\bibinfo {year} {1996})}\BibitemShut {NoStop}%
\bibitem [{\citenamefont {Giannozzi}\ \emph {et~al.}(2017)\citenamefont
  {Giannozzi}, \citenamefont {Andreussi}, \citenamefont {Brumme}, \citenamefont
  {Bunau}, \citenamefont {Nardelli}, \citenamefont {Calandra}, \citenamefont
  {Car}, \citenamefont {Cavazzoni}, \citenamefont {Ceresoli}, \citenamefont
  {Cococcioni}, \citenamefont {Colonna}, \citenamefont {Carnimeo},
  \citenamefont {Corso}, \citenamefont {de~Gironcoli}, \citenamefont {Delugas},
  \citenamefont {DiStasio}, \citenamefont {Ferretti}, \citenamefont {Floris},
  \citenamefont {Fratesi}, \citenamefont {Fugallo}, \citenamefont {Gebauer},
  \citenamefont {Gerstmann}, \citenamefont {Giustino}, \citenamefont {Gorni},
  \citenamefont {Jia}, \citenamefont {Kawamura}, \citenamefont {Ko},
  \citenamefont {Kokalj}, \citenamefont {K\"{u}\c{c}\"{u}kbenli}, \citenamefont
  {Lazzeri}, \citenamefont {Marsili}, \citenamefont {Marzari}, \citenamefont
  {Mauri}, \citenamefont {Nguyen}, \citenamefont {Nguyen}, \citenamefont {de-la
  Roza}, \citenamefont {Paulatto}, \citenamefont {Ponc\'{e}}, \citenamefont
  {Rocca}, \citenamefont {Sabatini}, \citenamefont {Santra}, \citenamefont
  {Schlipf}, \citenamefont {Seitsonen}, \citenamefont {Smogunov}, \citenamefont
  {Timrov}, \citenamefont {Thonhauser}, \citenamefont {Umari}, \citenamefont
  {Vast}, \citenamefont {Wu},\ and\ \citenamefont {Baroni}}]{Giannozzi}%
  \BibitemOpen
  \bibfield  {author} {\bibinfo {author} {\bibfnamefont {P.}~\bibnamefont
  {Giannozzi}}, \bibinfo {author} {\bibfnamefont {O.}~\bibnamefont
  {Andreussi}}, \bibinfo {author} {\bibfnamefont {T.}~\bibnamefont {Brumme}},
  \bibinfo {author} {\bibfnamefont {O.}~\bibnamefont {Bunau}}, \bibinfo
  {author} {\bibfnamefont {M.~B.}\ \bibnamefont {Nardelli}}, \bibinfo {author}
  {\bibfnamefont {M.}~\bibnamefont {Calandra}}, \bibinfo {author}
  {\bibfnamefont {R.}~\bibnamefont {Car}}, \bibinfo {author} {\bibfnamefont
  {C.}~\bibnamefont {Cavazzoni}}, \bibinfo {author} {\bibfnamefont
  {D.}~\bibnamefont {Ceresoli}}, \bibinfo {author} {\bibfnamefont
  {M.}~\bibnamefont {Cococcioni}}, \bibinfo {author} {\bibfnamefont
  {N.}~\bibnamefont {Colonna}}, \bibinfo {author} {\bibfnamefont
  {I.}~\bibnamefont {Carnimeo}}, \bibinfo {author} {\bibfnamefont {A.~D.}\
  \bibnamefont {Corso}}, \bibinfo {author} {\bibfnamefont {S.}~\bibnamefont
  {de~Gironcoli}}, \bibinfo {author} {\bibfnamefont {P.}~\bibnamefont
  {Delugas}}, \bibinfo {author} {\bibfnamefont {R.~A.}\ \bibnamefont
  {DiStasio}}, \bibinfo {author} {\bibfnamefont {A.}~\bibnamefont {Ferretti}},
  \bibinfo {author} {\bibfnamefont {A.}~\bibnamefont {Floris}}, \bibinfo
  {author} {\bibfnamefont {G.}~\bibnamefont {Fratesi}}, \bibinfo {author}
  {\bibfnamefont {G.}~\bibnamefont {Fugallo}}, \bibinfo {author} {\bibfnamefont
  {R.}~\bibnamefont {Gebauer}}, \bibinfo {author} {\bibfnamefont
  {U.}~\bibnamefont {Gerstmann}}, \bibinfo {author} {\bibfnamefont
  {F.}~\bibnamefont {Giustino}}, \bibinfo {author} {\bibfnamefont
  {T.}~\bibnamefont {Gorni}}, \bibinfo {author} {\bibfnamefont
  {J.}~\bibnamefont {Jia}}, \bibinfo {author} {\bibfnamefont {M.}~\bibnamefont
  {Kawamura}}, \bibinfo {author} {\bibfnamefont {H.-Y.}\ \bibnamefont {Ko}},
  \bibinfo {author} {\bibfnamefont {A.}~\bibnamefont {Kokalj}}, \bibinfo
  {author} {\bibfnamefont {E.}~\bibnamefont {K\"{u}\c{c}\"{u}kbenli}}, \bibinfo
  {author} {\bibfnamefont {M.}~\bibnamefont {Lazzeri}}, \bibinfo {author}
  {\bibfnamefont {M.}~\bibnamefont {Marsili}}, \bibinfo {author} {\bibfnamefont
  {N.}~\bibnamefont {Marzari}}, \bibinfo {author} {\bibfnamefont
  {F.}~\bibnamefont {Mauri}}, \bibinfo {author} {\bibfnamefont {N.~L.}\
  \bibnamefont {Nguyen}}, \bibinfo {author} {\bibfnamefont {H.-V.}\
  \bibnamefont {Nguyen}}, \bibinfo {author} {\bibfnamefont {A.~O.}\
  \bibnamefont {de-la Roza}}, \bibinfo {author} {\bibfnamefont
  {L.}~\bibnamefont {Paulatto}}, \bibinfo {author} {\bibfnamefont
  {S.}~\bibnamefont {Ponc\'{e}}}, \bibinfo {author} {\bibfnamefont
  {D.}~\bibnamefont {Rocca}}, \bibinfo {author} {\bibfnamefont
  {R.}~\bibnamefont {Sabatini}}, \bibinfo {author} {\bibfnamefont
  {B.}~\bibnamefont {Santra}}, \bibinfo {author} {\bibfnamefont
  {M.}~\bibnamefont {Schlipf}}, \bibinfo {author} {\bibfnamefont {A.~P.}\
  \bibnamefont {Seitsonen}}, \bibinfo {author} {\bibfnamefont {A.}~\bibnamefont
  {Smogunov}}, \bibinfo {author} {\bibfnamefont {I.}~\bibnamefont {Timrov}},
  \bibinfo {author} {\bibfnamefont {T.}~\bibnamefont {Thonhauser}}, \bibinfo
  {author} {\bibfnamefont {P.}~\bibnamefont {Umari}}, \bibinfo {author}
  {\bibfnamefont {N.}~\bibnamefont {Vast}}, \bibinfo {author} {\bibfnamefont
  {X.}~\bibnamefont {Wu}},\ and\ \bibinfo {author} {\bibfnamefont
  {S.}~\bibnamefont {Baroni}},\ }\href
  {https://doi.org/10.1088/1361-648X/aa8f79} {\bibfield  {journal} {\bibinfo
  {journal} {Journal of Physics: Condensed Matter}\ }\textbf {\bibinfo {volume}
  {29}},\ \bibinfo {pages} {465901} (\bibinfo {year} {2017})}\BibitemShut
  {NoStop}%
\bibitem [{\citenamefont {Hamann}(2013)}]{HamannPRB}%
  \BibitemOpen
  \bibfield  {author} {\bibinfo {author} {\bibfnamefont {D.~R.}\ \bibnamefont
  {Hamann}},\ }\href {https://doi.org/10.1103/PhysRevB.88.085117} {\bibfield
  {journal} {\bibinfo  {journal} {Phys. Rev. B}\ }\textbf {\bibinfo {volume}
  {88}},\ \bibinfo {pages} {085117} (\bibinfo {year} {2013})}\BibitemShut
  {NoStop}%
\bibitem [{\citenamefont {Schlipf}\ and\ \citenamefont {Gygi}(2015)}]{Schlipf}%
  \BibitemOpen
  \bibfield  {author} {\bibinfo {author} {\bibfnamefont {M.}~\bibnamefont
  {Schlipf}}\ and\ \bibinfo {author} {\bibfnamefont {F.}~\bibnamefont {Gygi}},\
  }\href {https://doi.org/https://doi.org/10.1016/j.cpc.2015.05.011} {\bibfield
   {journal} {\bibinfo  {journal} {Computer Physics Communications}\ }\textbf
  {\bibinfo {volume} {196}},\ \bibinfo {pages} {36} (\bibinfo {year}
  {2015})}\BibitemShut {NoStop}%
\bibitem [{\citenamefont {Marzari}\ and\ \citenamefont
  {Vanderbilt}(1997)}]{MarzariPRB}%
  \BibitemOpen
  \bibfield  {author} {\bibinfo {author} {\bibfnamefont {N.}~\bibnamefont
  {Marzari}}\ and\ \bibinfo {author} {\bibfnamefont {D.}~\bibnamefont
  {Vanderbilt}},\ }\href {https://doi.org/10.1103/PhysRevB.56.12847} {\bibfield
   {journal} {\bibinfo  {journal} {Phys. Rev. B}\ }\textbf {\bibinfo {volume}
  {56}},\ \bibinfo {pages} {12847} (\bibinfo {year} {1997})}\BibitemShut
  {NoStop}%
\bibitem [{\citenamefont {Souza}\ \emph {et~al.}(2001)\citenamefont {Souza},
  \citenamefont {Marzari},\ and\ \citenamefont {Vanderbilt}}]{SouzaPRB}%
  \BibitemOpen
  \bibfield  {author} {\bibinfo {author} {\bibfnamefont {I.}~\bibnamefont
  {Souza}}, \bibinfo {author} {\bibfnamefont {N.}~\bibnamefont {Marzari}},\
  and\ \bibinfo {author} {\bibfnamefont {D.}~\bibnamefont {Vanderbilt}},\
  }\href {https://doi.org/10.1103/PhysRevB.65.035109} {\bibfield  {journal}
  {\bibinfo  {journal} {Phys. Rev. B}\ }\textbf {\bibinfo {volume} {65}},\
  \bibinfo {pages} {035109} (\bibinfo {year} {2001})}\BibitemShut {NoStop}%
\bibitem [{\citenamefont {Seo}\ \emph {et~al.}(2015)\citenamefont {Seo},
  \citenamefont {Tsumuraya}, \citenamefont {Tsuchiizu}, \citenamefont
  {Miyazaki},\ and\ \citenamefont {Kato}}]{SeoJPSJ}%
  \BibitemOpen
  \bibfield  {author} {\bibinfo {author} {\bibfnamefont {H.}~\bibnamefont
  {Seo}}, \bibinfo {author} {\bibfnamefont {T.}~\bibnamefont {Tsumuraya}},
  \bibinfo {author} {\bibfnamefont {M.}~\bibnamefont {Tsuchiizu}}, \bibinfo
  {author} {\bibfnamefont {T.}~\bibnamefont {Miyazaki}},\ and\ \bibinfo
  {author} {\bibfnamefont {R.}~\bibnamefont {Kato}},\ }\href
  {https://doi.org/10.7566/jpsj.84.044716} {\bibfield  {journal} {\bibinfo
  {journal} {J. Phys. Soc. Jpn.}\ }\textbf {\bibinfo {volume} {84}},\ \bibinfo
  {pages} {044716} (\bibinfo {year} {2015})}\BibitemShut {NoStop}%
\bibitem [{\citenamefont {Fujiyama}\ and\ \citenamefont
  {Kato}(2019)}]{FujiyamaPRL}%
  \BibitemOpen
  \bibfield  {author} {\bibinfo {author} {\bibfnamefont {S.}~\bibnamefont
  {Fujiyama}}\ and\ \bibinfo {author} {\bibfnamefont {R.}~\bibnamefont
  {Kato}},\ }\href {https://doi.org/10.1103/PhysRevLett.122.147204} {\bibfield
  {journal} {\bibinfo  {journal} {Phys. Rev. Lett.}\ }\textbf {\bibinfo
  {volume} {122}},\ \bibinfo {pages} {147204} (\bibinfo {year}
  {2019})}\BibitemShut {NoStop}%
\bibitem [{\citenamefont {Miyazaki}\ and\ \citenamefont
  {Ohno}(2003)}]{MiyazakiPRB}%
  \BibitemOpen
  \bibfield  {author} {\bibinfo {author} {\bibfnamefont {T.}~\bibnamefont
  {Miyazaki}}\ and\ \bibinfo {author} {\bibfnamefont {T.}~\bibnamefont
  {Ohno}},\ }\href {https://doi.org/10.1103/PhysRevB.68.035116} {\bibfield
  {journal} {\bibinfo  {journal} {Phys. Rev. B}\ }\textbf {\bibinfo {volume}
  {68}},\ \bibinfo {pages} {035116} (\bibinfo {year} {2003})}\BibitemShut
  {NoStop}%
\bibitem [{\citenamefont {Coldea}\ \emph {et~al.}(2001)\citenamefont {Coldea},
  \citenamefont {Tennant}, \citenamefont {Tsvelik},\ and\ \citenamefont
  {Tylczynski}}]{ColdeaPRL}%
  \BibitemOpen
  \bibfield  {author} {\bibinfo {author} {\bibfnamefont {R.}~\bibnamefont
  {Coldea}}, \bibinfo {author} {\bibfnamefont {D.~A.}\ \bibnamefont {Tennant}},
  \bibinfo {author} {\bibfnamefont {A.~M.}\ \bibnamefont {Tsvelik}},\ and\
  \bibinfo {author} {\bibfnamefont {Z.}~\bibnamefont {Tylczynski}},\ }\href
  {http://dx.doi.org/10.1103/PhysRevLett.86.1335} {\bibfield  {journal}
  {\bibinfo  {journal} {Phys. Rev. Lett.}\ }\textbf {\bibinfo {volume} {86}},\
  \bibinfo {pages} {1335} (\bibinfo {year} {2001})}\BibitemShut {NoStop}%
\bibitem [{\citenamefont {Kohno}\ \emph {et~al.}(2007)\citenamefont {Kohno},
  \citenamefont {Starykh},\ and\ \citenamefont {Balents}}]{Kohno}%
  \BibitemOpen
  \bibfield  {author} {\bibinfo {author} {\bibfnamefont {M.}~\bibnamefont
  {Kohno}}, \bibinfo {author} {\bibfnamefont {O.~A.}\ \bibnamefont {Starykh}},\
  and\ \bibinfo {author} {\bibfnamefont {L.}~\bibnamefont {Balents}},\ }\href
  {http://dx.doi.org/10.1038/nphys749} {\bibfield  {journal} {\bibinfo
  {journal} {Nat. Phys.}\ }\textbf {\bibinfo {volume} {3}},\ \bibinfo {pages}
  {790} (\bibinfo {year} {2007})}\BibitemShut {NoStop}%
\bibitem [{\citenamefont {Starykh}\ and\ \citenamefont
  {Balents}(2007)}]{DimensionalReduction}%
  \BibitemOpen
  \bibfield  {author} {\bibinfo {author} {\bibfnamefont {O.~A.}\ \bibnamefont
  {Starykh}}\ and\ \bibinfo {author} {\bibfnamefont {L.}~\bibnamefont
  {Balents}},\ }\href@noop {} {\bibfield  {journal} {\bibinfo  {journal} {Phys.
  Rev. Lett.}\ }\textbf {\bibinfo {volume} {98}},\ \bibinfo {pages} {077205}
  (\bibinfo {year} {2007})}\BibitemShut {NoStop}%
\bibitem [{\citenamefont {Hayashi}\ and\ \citenamefont
  {Ogata}(2007)}]{Hayashi}%
  \BibitemOpen
  \bibfield  {author} {\bibinfo {author} {\bibfnamefont {Y.}~\bibnamefont
  {Hayashi}}\ and\ \bibinfo {author} {\bibfnamefont {M.}~\bibnamefont
  {Ogata}},\ }\href {http://dx.doi.org/10.1143/JPSJ.76.053705} {\bibfield
  {journal} {\bibinfo  {journal} {J. Phys. Soc. Jpn.}\ }\textbf {\bibinfo
  {volume} {76}},\ \bibinfo {pages} {053705} (\bibinfo {year}
  {2007})}\BibitemShut {NoStop}%
\bibitem [{\citenamefont {Yunoki}\ and\ \citenamefont
  {Sorella}(2006)}]{Yunoki}%
  \BibitemOpen
  \bibfield  {author} {\bibinfo {author} {\bibfnamefont {S.}~\bibnamefont
  {Yunoki}}\ and\ \bibinfo {author} {\bibfnamefont {S.}~\bibnamefont
  {Sorella}},\ }\href {https://doi.org/10.1103/PhysRevB.74.014408} {\bibfield
  {journal} {\bibinfo  {journal} {Phys. Rev. B}\ }\textbf {\bibinfo {volume}
  {74}},\ \bibinfo {pages} {014408} (\bibinfo {year} {2006})}\BibitemShut
  {NoStop}%
\end{thebibliography}

\end{document}